\documentclass{aa}  

\usepackage{natbib}
\usepackage{graphicx}
\usepackage{txfonts}
\usepackage{hyperref}
\usepackage{float}
\usepackage{ulem}
\usepackage{xcolor}

\bibpunct{(}{)}{;}{a}{}{,} % to follow the A&A style

\begin{document} 
   \title{ Tracing satellite planes in the Sculptor group:}
 \subtitle{I. Discovery of three faint dwarf galaxies around NGC\,253}  
   
   \titlerunning{Discovery of three satellites of NGC\,253}
    \authorrunning{Mart{\'\i}nez-Delgado et al.}

%   \subtitle{I. Overviewing the $\kappa$-mechanism}

   \author{David Mart{\'\i}nez-Delgado\inst{1}\thanks{Talentia Senior Fellow}, Dmitry Makarov$^{2}$, Behnam Javanmardi$^{3}$, Marcel S. Pawlowski$^{4}$, Lidia Makarova$^{2}$, Giuseppe Donatiello$^{5}$, Dustin Lang$^{6}$, Javier Román$^{1,7,8}$, Kathy Vivas$^{9}$, Julio A. Carballo-Bello$^{10}$}
   
\institute{
$^{1}$ Instituto de Astrofísica de Andalucía, CSIC, E-18080 Granada, Spain\\
$^{2}$ Special Astrophysical Observatory of the Russian Academy of Sciences, Nizhnij Arkhyz, 369167, Russia \\
$^{3}$ LESIA, Observatoire de Paris, Université PSL, CNRS, Sorbonne Université, Université de Paris, 5 place Jules Janssen, 92195 Meudon, France\\
$^{4}$ Leibniz-Institut f\"ur Astrophysik Potsdam (AIP), An der Sternwarte 16, D-14482 Potsdam, Germany\\
$^{5}$UAI -- Unione Astrofili Italiani /P.I. Sezione Nazionale di Ricerca Profondo Cielo, 72024 Oria, Italy \\
$^{6}$Perimeter Institute for Theoretical Physics, 31 Caroline St N, Waterloo, Canada\\
$^{7}$ Instituto de Astrof\'{\i}sica de Canarias, c/ V\'{\i}a L\'actea s/n, E-38205, La Laguna, Tenerife, Spain\\
$^{8}$ Departamento de Astrof\'{\i}sica, Universidad de La Laguna, E-38206, La Laguna, Tenerife, Spain\\
$^{9}$ Cerro Tololo Inter-American Observatory, NSF’s NOIRLab, Casilla 603, La Serena, Chile\\
$^{10}$ Instituto de Alta Investigaci\'on, Sede Esmeralda, Universidad de Tarapac\'a, Av. Luis Emilio Recabarren 2477, Iquique, Chile\\
}

\date{April 2021}

% \abstract{}{}{}{}{} 
% 5 {} token are mandatory
 
  \abstract
  {In the last years, a new generation of large-scale imaging surveys have probed for the first time wide field regions around some nearby galaxies at unprecedented low surface brightness regime ($\sim$ 28.0--29.0\,mag\,arcsec$^{-2}$). 
This offers a chance of discovering very faint dwarf satellites by means of systematic visual inspection of these public deep images.}
  {In this paper we report the first results of a systematic survey of faint dwarf spheroidal galaxies in the vicinity of the bright late-type spiral NGC\,253 galaxy by means of a visual inspection of the images taken by the Dark Energy Survey.}
   {We performed a new NGC\,253 satellite search using coadded image cutouts reprocessed in the DESI Legacy imagen surveys.
   We used \textsc{galfit} software for the photometric and structural properties of three dwarf galaxies.  }
  {Three new dwarf galaxies have been discovered in the vicinity of the brightest member of the Sculptor filament, the late-type spiral NGC\,253, located at a distance of 3.7~Mpc towards Anti-Virgo.
   We named them Do\,II, Do\,III and Do\,IV. Assuming they are companions of NGC\,253, their total absolute $V$-magnitudes fall in the $-7$ to $-9$\,mag range, which is typical for dwarf satellites in the local Universe.The central surface brightness tend to be extremely low for all the discovered dwarfs and fall roughly in the range of 25--26\,mag\,arcsec$^{-2}$ in $g$-band. Using known data on distances and velocities of galaxies, we estimate the total virial mass of the NGC\,253 group to be $8\times10^{11}M_{\sun}$, which gives the virial radius $R_{200} = 186$\,kpc and the turn-around radius of 706\,kpc.
   We also discuss the possible existence of a spatially flattened and velocity-correlated satellite system around NGC\,253. This large-scale structure is orientated almost edge-on to line of sight.
   The possible plane of satellites is only 31\,kpc thick with the minor-to-major axis ratio of 0.14. 
   Four out of five galaxies with measured velocities follow a common velocity trend similar to those observed in the planes of satellites around the Andromeda and Centaurus\,A galaxies.
   However, the small number of galaxies with known velocities prevents to reach a definitive conclusion about the formation scenario of the structure and its possible relation to the surrounding cosmic web.  }
   {}


\maketitle

%
%------------------------------------------------------------------
%%%%%%%%%%%%%%%%% BODY OF PAPER %%%%%%%%%%%%%%%%%%
%[\textcolor{blue}{Behnam: A general comment, please add all the references to the .bib file. It is significantly easier to use BibTeX than to write the individual references.}]
\section{Introduction}

Dwarf galaxies are rich laboratories for studying stellar populations \citep{Tolstoy2000,Grebel2005}, galaxy formation scenarios, cosmological models, and gravitational theories \citep{Kroupa2018}. The observations of dwarf galaxies in the Local Group (LG) has faced the standard Lambda-Cold-Dark-Matter ($\Lambda$CDM) cosmological model with a number of challenges \citep[see e.g.][for a review]{Bullock2017}. In fact, some recent observational studies indicate that the challenges to $\Lambda$CDM appear not to be limited to only the LG.
\citet{Chiboucas_2013} and \citet{Mueller2018} found discs of satellites around M\,81 and the Centaurus\,A galaxy similar to those rotating around the Milky Way \citep{PawlowskiKroupa2013} and M\,31 \citep{Ibata2013};
\citet{Smercina2018} and \citet{Bennet2019} reported a \textit{missing satellites} problem in the M\,94 and M\,101 groups;
and \citet{Javanmardi2020} found an unexpected correlation between the number of satellites and the bulge-to-total baryonic mass ratio extending beyond the LG. 
These findings require further deep surveys of faint dwarf galaxies around as many massive galaxies in the local volume as possible.

$\Lambda$CDM simulations predict that subhalos around more massive host halos, such as that expected to surround the Milky Way, are mostly randomly isotropically distributed and have largely uncorrelated relative velocities. 
While some anisotropy in distribution and some coherence in motion is expected -- induced by the preferential accretion of subhalos along cosmic filaments and in small groups \citep{Libeskind2011, Pawlowski2012, Shao2018} -- an overall close-to-isotropic distribution of satellite systems is a robust prediction of the underlying model. 
This is because the positions and motions of satellites on scales of 100's of kpc are not strongly affected by the intricacies of baryonic physics, or the minutiae of how these are implemented in cosmological simulations.
Testing this prediction for the phase-space distribution of satellite galaxies has revealed a serious challenge to $\Lambda$CDM in the LG: the observed satellite galaxy systems of the Milky Way \citep{PawlowskiKroupa2013} and M\,31 \citep{Ibata2013} display flattened distributions, whose kinematics indicate a preference of satellites to co-orbit along these structures. 
Recent proper motion data indicate that a substantial fraction of the Milky Way satellites \citep{Fritz2018, Li2021EDR3}, as well as at least two of M\,31's on-plane satellites \citep{Sohn2020}, indeed orbit along these planes of satellite galaxies. 
Simulated systems with similar degrees of coherence are exceedingly rare in cosmological simulations \citep{Ibata2014}. 
For a review, see \citet{Pawlowski2018}.

Some similar satellite alignments have been identified outside the LG. 
The most prominent to date is a flattened distribution of satellites, close to an edge-on orientation, identified around Centaurus\,A by \citet{Tully2015}. 
To this, \citet{Mueller2018} added evidence for line-of-sight velocity kinematics being consistent with a rotating satellite plane, a finding recently confirmed when additional spectroscopic velocity measurements brought the number of kinematically correlated satellites to 21 out of 28 \citep{Mueller2021}. 
Some additional hints at  similar structures exist around other host galaxies such as M\,81 \citep{Chiboucas_2013}, as well as in a statistical analysis of satellite galaxy pairs \citep{Ibata2014}. 
Nevertheless, the cosmological challenge posted by the planes of satellite galaxies rests so far on only a small number of studied systems.  For a better and statistically more reliable understanding, we need to aim for an as complete census of satellites around other nearby hosts as possible. Systems in the vicinity of the LG are accessible for follow-up measurements of photometric distances and line-of-sight velocities of its members.
It allows one to understand their the three-dimensional structure and kinematics.
Assembling a more complete picture of nearby satellite galaxy systems is thus fundamental to confirm some proposed formation scenarios for the observed alignments from an observational perspective, an urgently needed addition to complement the (so far) mostly simulation-driven debate.

Ultra-deep imaging in wide sky areas with amateur telescopes \citep{2016A&A...588A..89J,2016MNRAS.457L.103R,2020AN....341.1037K} can also help to complete the census of these hitherto unknown low surface brightness (LSB) galaxies. 
In the last one and half decades, the discoveries of dwarf satellites in the LG have been made using stellar density maps of resolved stars, counted in selected areas of the color-magnitude diagrams (CMDs) from large scale surveys such as the Sloan Digital Sky Survey (SDSS) \citep[e.g.,][]{2005ApJ...626L..85W, 2006ApJ...650L..41Z}, Pan-STARRs \citep[e.g.,][]{2015ApJ...813...44L}, Dark Energy Survey (DES) \citep[]{DES16, 2015ApJ...807...50B} and, more recently, the DECam Local Volume Exploration Survey (DELVE) \citep[]{Drlica21,2020ApJ...890..136M}, Hyper Suprime-Cam Subaru Strategic Program (HSC-SSP) \citep[]{homma19} and the DESI Imaging Legacy surveys \citep[]{2021arXiv210403859M}.
%\sout{
%In the case of the Milky Way, the main tracers to find these diffuse systems are blue stars 2--3~mag %fainter than the main sequence (MS) turnoff of the old stellar population. 
%However,the relatively shallow photometry of these surveys (with a $g$-band limiting magnitude for point sources of $\sim$21.5) and the significant contamination by foreground stars and distant blue galaxies, make it  difficult to complete the census of faint  dwarf companions at distances larger than 100\,kpc using these data. }

In the case of the Milky Way, these diffuse systems are usually found as over-densities of old stellar populations in certain regions of the sky. Contamination by foreground stars and background galaxies makes this task challenging for diffuse systems. 
However, modern surveys are now routinely finding dwarf galaxies within the virial radius of the Milky Way ($\sim 300$ kpc) \citep[see the recent review by][and references therein]{2019ARA&A..57..375S}. 
The larger distance of the Andromeda galaxy makes it prohibitive  to resolve their stellar halo stars fainter than the red clump/horizontal branch level \citet[e.g., PAndAS photometry:][]{2019AJ....157..168D}.
Thus, the M\,31 satellite population  can only be traced by observations of the less numerous, bright red-giant branch (RGB) stars. 
This means that the dwarf galaxy hunting in Andromeda is certainly biased  due to the lack of enough stellar tracers for dwarf galaxies with absolute magnitude fainter than about $-6$\,mag, which are barely resolved into stars with ground-based telescopes. 
In the last years, new generation large-scale imaging surveys \citep[e.g., the DESI imaging Legacy surveys:][]{2019AJ....157..168D}  have covered for the first time wide field regions around nearby galaxies at unprecedented low surface brightness regime ($\sim$ 28.0--28.5\,mag\,arcsec$^{-2}$). 
This offers the possibility of detecting hitherto diffuse dwarf satellites by means of systematic searches based on visual inspection of these public images \citep[e.g.,][]{2021arXiv210403859M} or on automatic detection algorithms \citep[e.g.,][]{2018ApJ...857..104G, 2021ApJS..252...18T, 2021OJAp....4E...3M, 2021MNRAS.500.2049P, 2021A&A...645A.107H}.

NGC\,253 is one of the closest spirals behind M\,31 and thus the natural place to dig for LSB dwarf galaxies that could provide new insights on the presence satellites planes around nearby galaxies outside the LG.  
At a distance of $\sim 3.7$\,Mpc, this galaxy in the Sculptor group has been explored for satellite galaxies in the past \citep{1997AJ....114.1313C,1998A&AS..127..409K,1998AJ....116.2873J,2000A&AS..146..359K,2000AJ....119..593J,2014ApJ...793L...7S,2016ApJ...816L...5T}.
In this paper, we present the discovery of three LSB dwarf galaxies near NGC\,253 by a visual inspection of the DES imaging data. Using an updated census of low-mass systems around this galaxy, we address for the first time the issue of the existence of a spatially flattened and velocity-correlated dwarf galaxy system around NGC\,253.

\begin{figure*}
   \begin{center}
	\includegraphics[width=0.90 \textwidth]{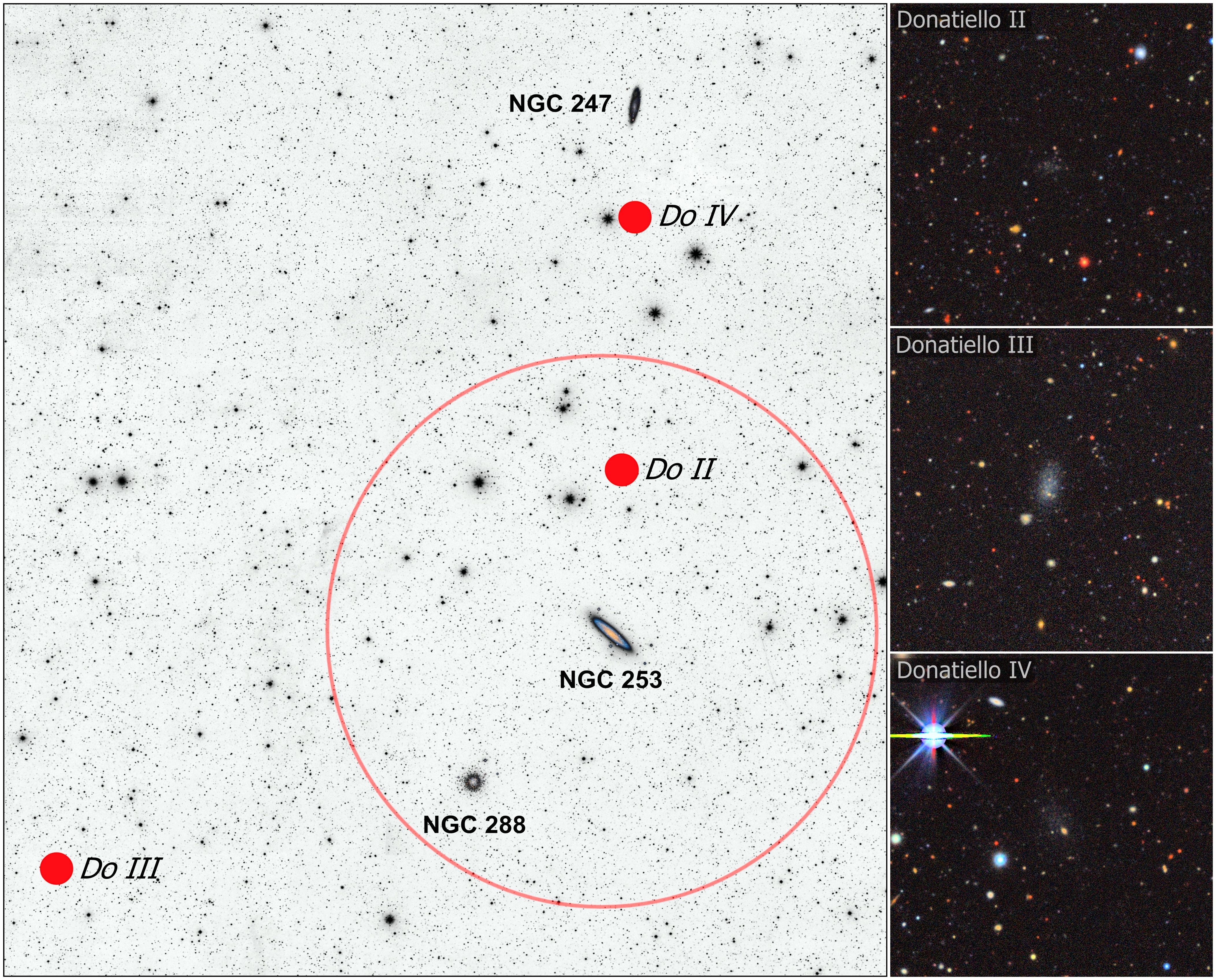}
    \caption{ ({\it Left panel}: The position of the three dwarf galaxies (solid red circles) reported in this study with respect to the spiral  NGC\,253. The red circular line corresponds to the area explored by the PISCes survey \citep{2016ApJ...816L...5T} extending up to $\sim$ 150 kpc from the center of NGC\,253.The total field-of-view of this image is 450\arcmin $\times$ 480\arcmin. {\it Left panel}: Full color version of the image cutouts obtained with {\it legacypipe} for Do~II, Do~III and Do~IV. North is up and East is left. The field of view of all these image cutouts is $3.3\arcmin \times 3.3\arcmin$.}
    \label{fig-map-ngc253}
     \end{center}
\end{figure*}

\section{SEARCHING STRATEGY AND DATA ANALYSIS}

\subsection{Searching strategy }

The dwarf galaxy candidates reported in this paper were found by the amateur astronomer Giuseppe Donatiello by visual inspection of the Dark Energy Camera \citep[DECam:][]{2015AJ....150..150F} images of the Scuptor group of galaxies  available from the DESI Legacy Imaging Surveys \citep{2019AJ....157..168D}. 
A total number of 13 candidates were detected in an total explored area of $15\times 10$ degrees. For this paper, we focus on only three candidates with clear evidence of being partially resolved into stars in the vecinity of NGC\,253, as shown in Fig.~\ref{fig-map-ngc253} (left panel).The typical angular resolution of these data (estimated from the seeing of the images) is $\sim$ 0.9\arcsec. We name them Donatiello~II (Do\,II), Donatiello~III (Do\,III) and Donatiello IV (Do\,IV), which position coordinates are given in Table~\ref{tab:params}.  These galaxies have not been detected in the automatic search of diffuse stellar systems in the DES data undertaking by \citet{2021ApJS..252...18T}.

\subsection{Image cutout data}

The DESI Legacy Imaging Surveys compile optical data in three optical bands ($g$, $r$ and $z$) coupled with all-sky infrared imaging from the Wide-field Infrared Survey Explorer (WISE) \citep{wright2010,meisner2019} and obtained by three different imaging projects on three different telescopes: The DECam Legacy Survey (DECaLS), the Beijing-Arizona Sky Survey (BASS) and the Mayall $z$-band Legacy Survey (MzLS) \citep{2019ApJS..245....4Z, 2019AJ....157..168D}. The DESI Legacy Imaging Surveys data releases also include re-reduced public DECam data from the DES \citep[][]{2018ApJS..239...18A}. 

Image cutouts centered on each satellite candidate were subsequently obtained by coadding images of these systems taken by the DES \cite[]{DES16} using the DECam. These data were reprocessed using the \textsc{legacypipe} software of the DESI Legacy imaging surveys \citep[see, for example, Fig. 2 in ][]{2021arXiv210406071M}. In short, each image is astrometrically calibrated to Gaia-DR2 and photometrically calibrated to the Pan-STARRS PS1 survey, and then resampled to a common pixel grid and summed with inverse-variance weighting.  Fig.~\ref{fig-map-ngc253} (right panel) shows the resulting coadded image cutouts of the three dwarf galaxies, which  include approximately 8 exposures in each of $g$, $r$, and $z$-bands. We have calculated the nominal depths of the images in the NGC~253 region following the approach by \citet{2020A&A...644A..42R}, appendix A. The calculated surface brightness limits are $\mu$= 29.3, 29.0 and 27.7 mag arcsec$^{-2}$ for the $g$, $r$ and $z$ bands respectively measured as 3$\sigma$ in 10$\times$10 arcsec boxes.

%The model is generated by forward-modeling simple elliptical galaxy profiles that are convolved by the PSF of each individual image.  The choice between different morphological types (point source, round exponential, elliptical exponential or de Vaucouleurs, or generic Sersic profile) is made on the basis of best-fit chi-squared values plus complexity priors.

%\begin{figure*}
%	\includegraphics[width=0.8\textwidth]{Panel1_Abril30.jpg}
%    \caption{Comparison with other facilities}
%    \label{fig-cmd}
%\end{figure*}

\begin{table}
\centering
      \caption{The \textsc{galfit} results for the coordinate of the center of the galaxy, the Sersic index (n), Axis ratio (b/a),  Position Angle (PA), and effective radius ($R_\mathrm{eff}$). The $V$-band magnitude is estimated adopting the transformation equation provided by \citet{Jester2005}. The physical parameters are calculated by adopting a distance modulus of $27.84\pm0.02$ mag \citep{2009AJ....138..332J} for NGC\,253.}
\begin{tabular}{lr@{ $\pm$ }lr@{ $\pm$ }lr@{ $\pm$ }l}
\hline\hline
                & \multicolumn{2}{c}{Do\,II}    & \multicolumn{2}{c}{Do\,III}   & \multicolumn{2}{c}{Do\,IV}\\
\hline
RA (J2000)      & \multicolumn{2}{r}{00 47 07.15}             
                                                & \multicolumn{2}{r}{01 09 24.55}             
                                                                                & \multicolumn{2}{r}{00 47 02.95} \\
Dec (J2000)     & \multicolumn{2}{r}{$-$23 57 20.9\phantom{0}}
                                                & \multicolumn{2}{r}{$-$27 20 49.6\phantom{0}}
                                                                                & \multicolumn{2}{r}{$-$21 40 51.6\phantom{0}} \\
n               & 0.62      & 0.12              & 0.56      & 0.03              & 0.85      & 0.12 \\
b/a             & 0.61      & 0.04              & 0.59      & 0.01              & 0.60      & 0.03 \\
PA (deg)        & $-89.1$   & 5.2               & $-10.4$   & 1.6               & +24.0     & 3.7 \\
$R_\mathrm{eff}$ (arcsec)   
                & 5.51      & 0.37              & 8.46      & 0.17              & 10.21     & 0.65 \\
$V$ (mag)       & 20.85     & 0.20              & 18.75     & 0.09              & 19.99     & 0.15 \\
$M_V$ (mag)     & $-7.04$   & 0.20              & $-9.13$   & 0.09              & $-7.89$   & 0.15 \\
$R_\mathrm{eff}$ (pc)       
                & 99        & 7                 & 152       & 3                 & 183       & 12 \\
%RA, DEC (deg)       & (11.7798, -23.9558)   & (17.3523, -27.3471)   & (11.7623, -21.6810) \\
%J2000                   & 004707.15$-$235720.9  & 010924.55$-$272049.6  & 004702.95$-$214051.6 \\
%PA (deg)           & $-89.11\pm5.20$       & $-10.36\pm1.61$       & $23.98\pm3.69$\\
%$V$ (mag)               & $20.72 \pm 0.54$      & $18.74\pm0.40$        & $19.96\pm 0.62$\\ % wrong errors
%$M_V$ (mag)         & $ -7.12 \pm 0.54$     & $-9.10\pm0.41$        & $-7.88\pm0.62$ \\ % not corrected for extinction
%$V$ (mag)       & 20.72     & 0.24              & 18.74     & 0.09              & 19.96     & 0.16 \\ % original data
%$M_V$ (mag)     & $-7.17$   & 0.24              & $-9.14$   & 0.09              & $-7.92$   & 0.17 \\ % original data
\hline\hline
\end{tabular}
\label{tab:params}
\end{table}

\subsection{ Photometry and Structural Properties}

\begin{figure*}
    \centering
    \includegraphics[width=\textwidth]{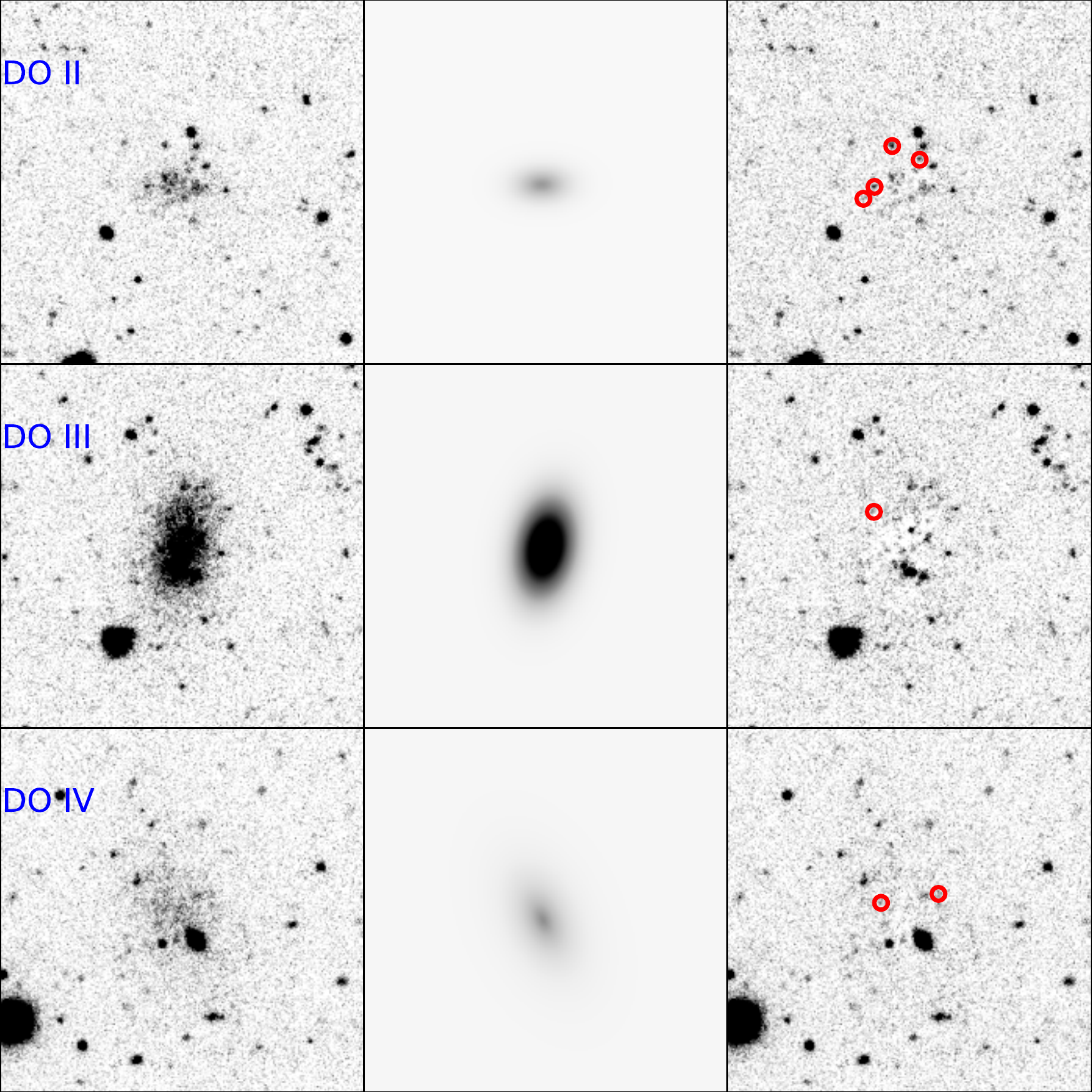}
    \caption{ From left to right: the dwarf galaxies in $r$-band, the \textsc{galfit} models, and the residual images after subtracting the model. The image scales and min max are the same for all the panels. The same procedure is done for modeling the $g$ and $z$-band images. We mark the position of a few partially resolved resources in the galaxies that are visible in the residual images with red circles.}
    \label{fig:GALFIT_result}
\end{figure*}

We use \textsc{galfit} software \citep{Peng2002} for determining the photometry and structural properties of dwarf galaxies following a similar approach to that described in \cite{2016A&A...588A..89J}. First, the images of the three $g$, $r$, and $z$-bands are combined to increase the signal-to-noise and to perform the first \textsc{galfit} modelling. The center of the galaxy, Sersic index, axis ratio, position angle, and the effective radius are obtained from the combined images and the results are listed in Table~\ref{tab:params}. In the next step, to stabilize the fitting procedure, these parameters are kept fixed to their best values while fitting the individual band images for measuring the magnitude and surface brightness which are listed in Table~\ref{tab:mags}. 
Fig.~\ref{fig:GALFIT_result} shows the results of \textsc{galfit} modeling in $r$-band. The modeling for $g$ and $z$-bands yields similar visual results hence are not shown here.

A careful inspection of the residual images (which is the result of subtracting the \textsc{galfit} model from the data) shows a number of unresolved sources which could belong to the dwarf galaxies. Among these unresolved sources, we mark those that pass the conditions $r>24.0$ and $0.4<g-r<1.2$ by red circles. 
%\textcolor{red}{
%\sout{
% By adding the fluxes of these marked sources to the total fluxes obtained via \textsc{galfit} modeling, we obtain more accurate estimates of the total magnitudes, $m_\mathrm{Tot}$, listed in the last row of Table~\ref{tab:mags}. The change in the magnitude in all the three bands are smaller than the final total uncertainties. 
%}
%}

%In addition, we made a second fitting with IMFIT method (Erwin 2014) of the galaxy, masking the brightest stars clearly not belonging to the galaxy. The photometry is consistent with the results using \textsc{galfit}, and the results are also shown in Table~\ref{tab:mags}. 

\begin{table*}
\caption{
First three rows: The \textsc{galfit} results for the integrated magnitude, $m_{G}$, the central surface brightness $\mu_{0,G}$ and the surface brightness at effective radius $\mu_{e,G}$ both in mag\,arcsec$^{-2}$ for $g$, $r$, and $z$-bands. 
}    
\centering
\small
\setlength{\tabcolsep}{3pt}
\begin{tabular}{lp{0pt}c@{\;\;}c@{\;\;}cp{0pt}c@{\;\;}c@{\;\;}cp{0pt}c@{\;\;}c@{\;\;}c}
\hline\hline
&&  \multicolumn{3}{c}{Do\,II} && \multicolumn{3}{c}{Do\,III} && \multicolumn{3}{c}{Do\,IV} \\
%\hline
&& $g$ & $r$ & $z$ && $g$ & $r$ & $z$ && $g$ & $r$ & $z$ \\
%\hline
\cline{3-5} \cline{7-9} \cline{11-13} 
$m_{G}$             &   & $21.01\pm0.15$ & $20.75\pm0.15$ & $20.51\pm0.15$ &    & $19.13\pm0.07$ & $18.50\pm0.05$ & $18.26\pm0.05$ &    & $20.37\pm0.12$& $19.74\pm0.10$ & $19.53\pm0.12$ \\
$\mu_{0,G}$         &   & $25.66\pm0.15$ & $25.40\pm0.15$ & $25.16\pm0.15$ &    & $24.76\pm0.07$ & $24.13\pm0.05$ & $23.89\pm0.05$ &    & $25.97\pm0.12$ & $25.34\pm0.10$ & $25.13\pm0.12$\\
$\mu_{e,G}$         &   & $26.66\pm0.15$ & $26.40\pm0.15$ & $26.16\pm0.15$ &    & $25.64\pm0.07$ & $25.00\pm0.05$ & $24.77\pm0.05$ &    & $27.47\pm0.12$ & $26.84\pm0.10$ & $26.64\pm0.12$\\ %[2pt]
%         \hline
%         $m_\mathrm{IM}$ & $20.77$ & $20.34$ & $20.25$  &  & & \\
%         \hline
%$m_\mathrm{Tot}$    &   & $20.91\pm0.42$ & $20.61\pm0.34$ & $20.29\pm0.47$ &    & $19.12\pm0.34$ & $18.49\pm0.22$ & $18.24\pm0.19$ &    & $20.35\pm0.56$	& $19.71\pm0.26$ & $19.49\pm0.73$ \\ % wrong errors
%$m_\mathrm{Tot}$    &   & $20.91\pm0.17$ & $20.61\pm0.17$ & $20.29\pm0.20$ &    & $19.12\pm0.07$ & $18.49\pm0.05$ & $18.24\pm0.05$ &    & $20.35\pm0.13$	& $19.71\pm0.10$ & $19.49\pm0.14$ \\
\hline\hline
\end{tabular}
\label{tab:mags}
\end{table*}

Using the $m_\mathrm{Tot}$ values and adopting the transformation equation provided by \citet{Jester2005}\footnote{\url{https://www.sdss.org/dr12/algorithms/sdssubvritransform}} $V = g - 0.59\times(g-r) - 0.01$, we obtain the $V$-band magnitudes. These results are also listed in Table~\ref{tab:params}. Table~\ref{tab:mags} shows as the surface brightness of these three stellar system is lower than 25 mag/arcsec$^2$) and thus cannot be detected in the previous photographic and CCD images from large-scale surveys like the POSS-II or PanSTARRs.

%\begin{figure}
%	\includegraphics[width=0.9\columnwidth]{CMD_June2.jpg}
%    \caption{Resolved stars within the main body of Do~II overlapping the CMD of SCL-MM-DW2 (Toloba et al. 2015)}
%    \label{fig-cmd}
%end{figure}

\section{DISCUSSION}

%\begin{figure*}
%	\includegraphics[width=1.0\textwidth]{Figura3_May21.jpg}
%    \caption{Comparison with other NGC\,253 satellites for checking level of resolution.}
%    \label{fig-cmd}
%\end{figure*}

\begin{figure*}
	\includegraphics[width=\textwidth]{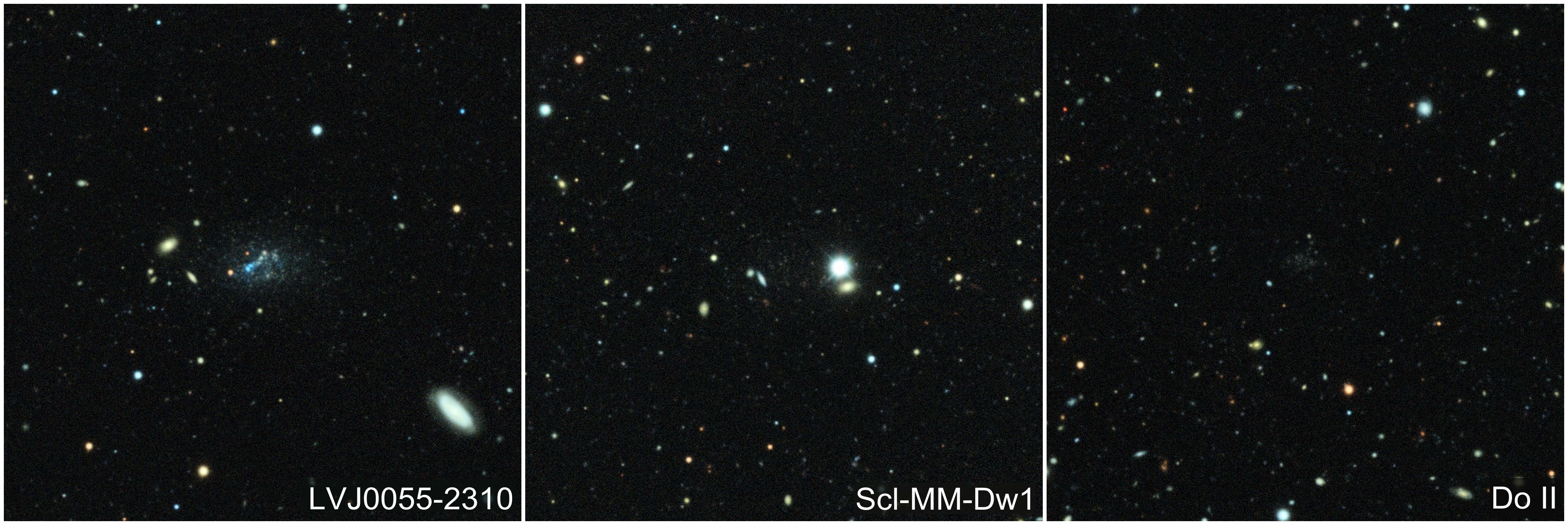}
    \caption{
    Comparison of the degree of resolution into stars of Do~II in the DES images 
    with the data from the same survey for two confirmed satellites of NGC\,253 satellites 
    situated at similar distance (see Table~\ref{tab:n253group}): LVJ005$-$2310 and Scl--MM--Dw1. 
    The field-of-view of these image cutouts are $4\arcmin \times 4\arcmin$. 
    North is up and East is left.
    }
    \label{fig:resolution}
\end{figure*}

\begin{figure}
    \centering
    \includegraphics[scale=0.5]{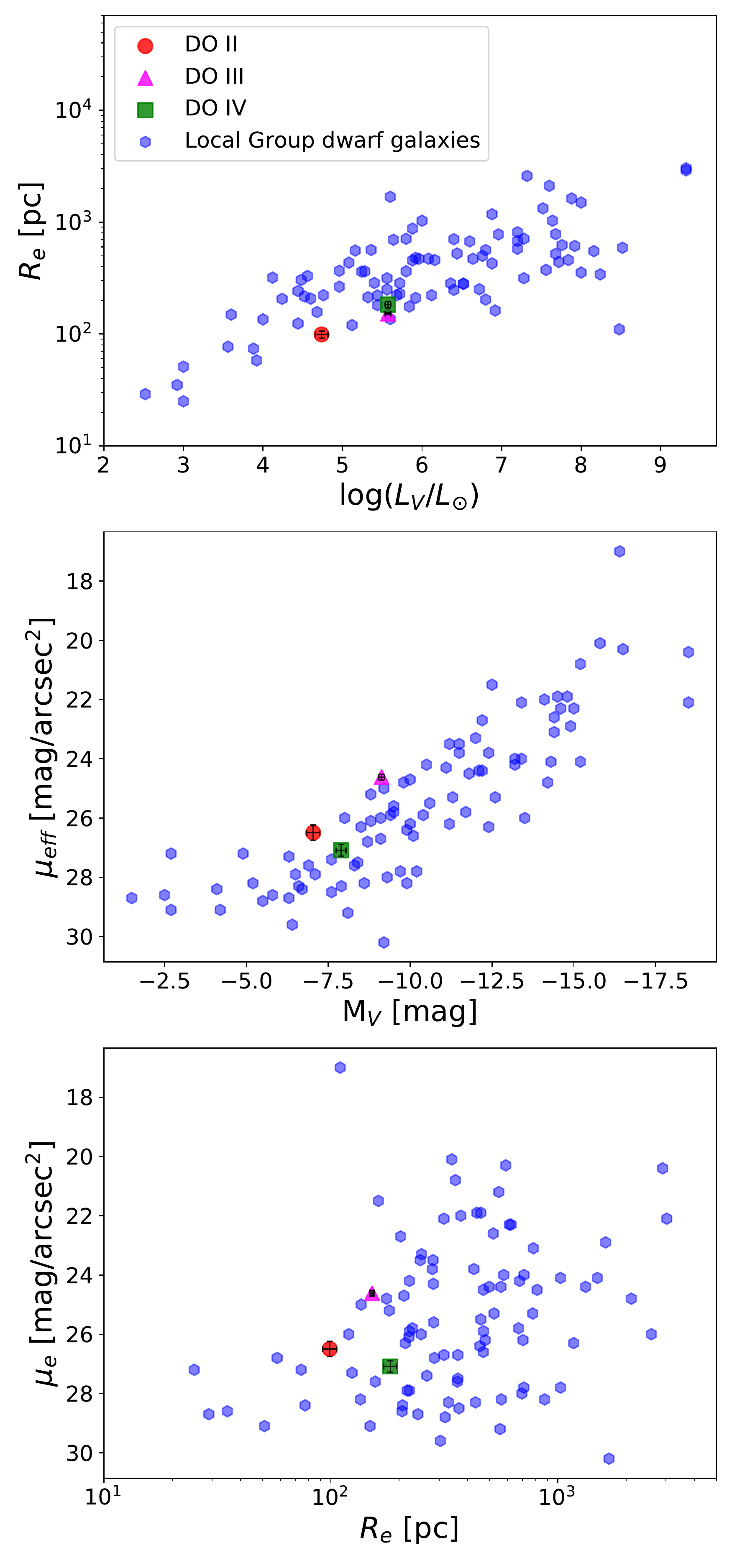}
    \caption{Comparison of the photometric and structural parameters of the three dwarf galaxies reported in this study with those for the Milky Way and M\,31 companions.}
    \label{fig:LGcomparison}
\end{figure}

\subsection{ Distances}

The DESI Legacy surveys images are not deep enough to build up a color-magnitude diagram of resolved stellar populations in these dwarf galaxies. Thus, it is not possible to determine their distance from the brightness of the tip of the RGB.  Given their close angular distance to NGC\,253, we assume that the discovered galaxies are at the same distance as NGC\,253, following a similar approach to that used in other recent wide-field surveys of dwarf satellite systems in the local volume \citep{2016A&A...588A..89J, 2020ApJ...891..144C}. This is also supported by the degree of resolution into stars displayed for these new objects when it is compared with the images of some confirmed dwarf companions of NGC\,253 obtained in the same survey (see Fig. \ref{fig:LGcomparison}). In addition, our hypothesis is consistent with the results from a standard approach used to establish membership based on the comparison of morphological and photometric properties with those from known dwarf galaxies \citep[see for instance][]{2018A&A...615A.105M}. In Table \ref{tab:params} we list the effective radius (in $pc$) and absolute magnitude, $M_V$ (in mag), 
by assuming a distance modulus of $27.84\pm0.02$ mag \citep{2009AJ....138..332J} for NGC\,253. 
Fig.~\ref{fig:resolution} shows the distribution of these three dwarf galaxies 
in the $R_e$ vs. $L$, $\mu_e$ vs. $R_e$, and $\mu_e$ vs. $M_V$ planes. 
The known dwarf galaxies of the Milky Way and Andromeda \citep{2012AJ....144....4M} are also shown for comparison. 
As can be seen, the three candidates can be characterized by similar properties as those of the LG dwarf galaxies,
supporting our assumption that they could be located at a similar distance like that of NGC\,253. However, with the present ground-based data, 
we cannot completely reject that some of these new dwarfs may be background galaxies 
projected onto the sky region of the Sculptor group.

\subsection{The NGC\,253 group of galaxies}

\begin{figure}
\centering
\includegraphics[width=0.49\textwidth]{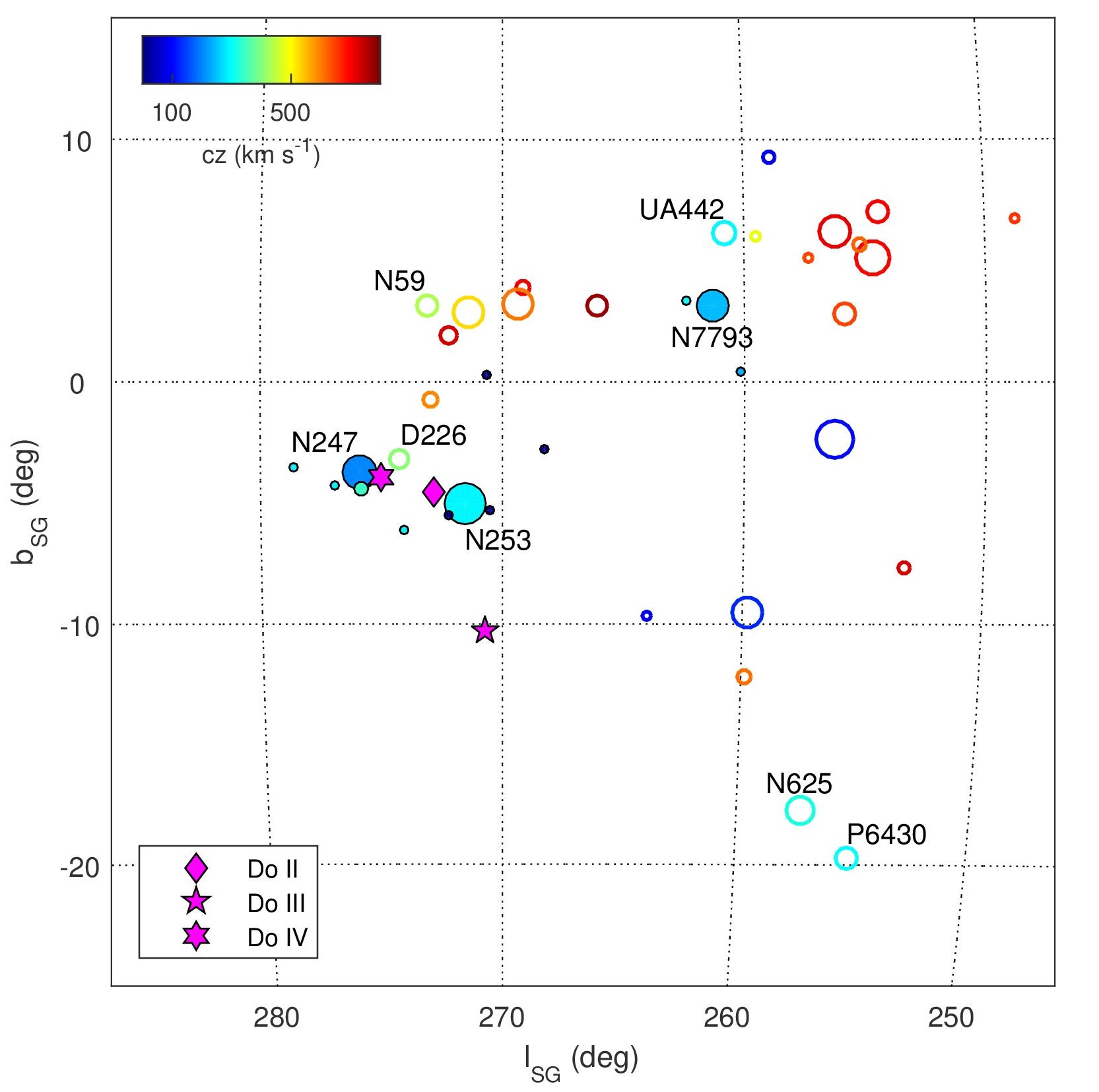}
\caption{
Distribution of galaxies around NGC\,253 in a $40\degr\times40\degr$ field in the supergalactic coordinates.
The members of the NGC\,253 group are shown by filled colored circles.
The background (reddish) and foreground (bluish) galaxies are marked with open circles.
The redshift is color-coded accordingly the color schema.
Newly discovered objects are shown by magenta polygons.
}
\label{fig:n253sky}
\end{figure}

Three new dwarf galaxies have been discovered in the vicinity of the bright late-type spiral NGC\,253.
This galaxy is the brightest member of the Sculptor filament \citep{1998AJ....116.2873J,2003A&A...404...93K}
located in the Local Supercluster plane almost in the Anti-Virgo direction.
The distribution of galaxies over the sky in the Sculptor filament is shown in the Fig.~\ref{fig:n253sky}.
This region has been intensively studied in a number of \ion{H}{I} surveys \citep[see for instance][]{2005AJ....130.2058B,2017MNRAS.472.4832W,2018MNRAS.478.1611K}.
The systematic searches of dwarf galaxies \citep{1997AJ....114.1313C,1998A&AS..127..409K,1998AJ....116.2873J,2000A&AS..146..359K,2000AJ....119..593J,2014ApJ...793L...7S}
have revealed a population of galaxies of very low luminosity down to $M_V\approx-10$\,mag.
High-precision distances using the tip of the red giant branch (TRGB) were measured in the works by \citet{2003A&A...404...93K,2003AJ....126.2806C,2005ApJ...633..810M,2006ApJ...641..822D,2006AJ....131.1361K,2008ApJ...686L..75M,2009AJ....138..332J,Dalcanton2009,2009AJ....137.4361D,2011ApJS..195...18R,2013A&A...549A..47L,2014ApJ...793L...7S,2016ApJ...816L...5T,2021arXiv210211354K}.

\begin{figure*}
\centering
\includegraphics[width=\textwidth]{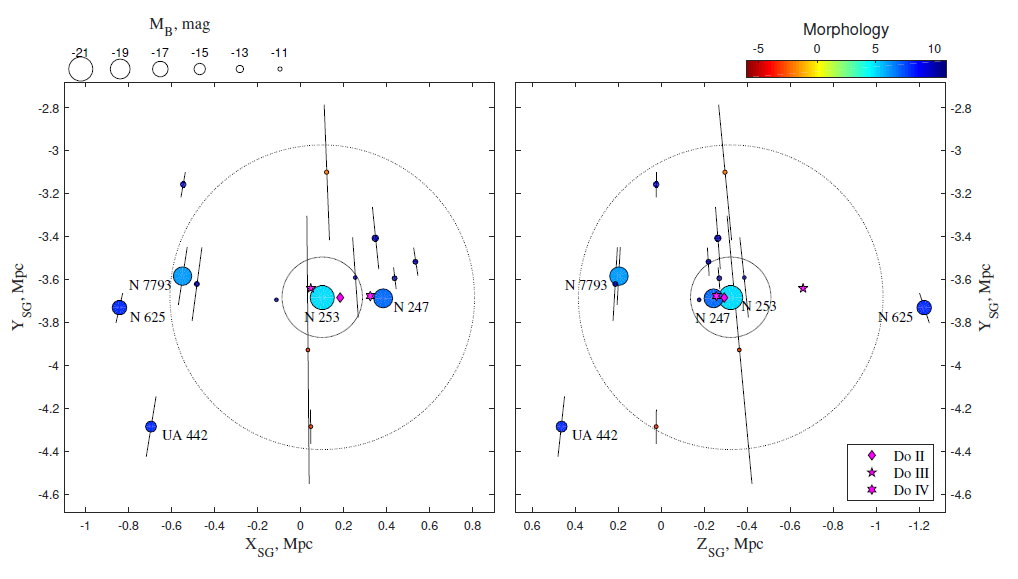}
\caption{
The map of galaxies in the vicinity of NGC\,253 in the supergalactic coordinates.
The left-hand panel presents the XY-projection on the supergalactic plane, 
while the right-hand panel shows the edge-on view of the Local Supercluster ‘pancake’.
The size and color of the dots correspond to the absolute $B$-band magnitude 
and morphology of galaxies, according to the schemes.
The line segments indicate uncertainties in the distance estimations.
Big circles around NGC\,253 mark the virial zone, $R_{200}\approx190$\,kpc,
and the zero velocity surface, $R_0\approx700$\,kpc.
The magenta symbols indicate the expected position of the discovered galaxies.
}
\label{fig:n253xyz}
\end{figure*}

\begin{table*}
\caption{Galaxies in the vicinity of NGC\,253.The columns contain: 
1) galaxy name;
2) morphological type;
3) $V_h$ -- heliocentric velocity in km\,s$^{-1}$ with its error;
4) $V_\mathrm{LG}$ -- radial velocity with respect to the LG centroid \citep{1996AJ....111..794K};
5) D$_\mathrm{TRGB}$ -- TRGB-distance in Mpc with a corresponding error;
6) $M_B$ -- $B$-band absolute magnitude of the galaxy;
7) TI -- `tidal index' indicating the value of the tidal influence from the main disturber \citep{1999IAUS..186..109K,2013AJ....145..101K};
8) $\Theta_\mathrm{N253}$ -- projected separation from NGC\,253 in degrees;
9) $R_\perp$ -- projected distance to NGC\,253 in kpc;
10) $R_\mathrm{3D}$ -- spatial distance to NGC\,253 in kpc. }
\label{tab:n253group}
\centering
\begin{tabular}{lllclrrrrr}
\hline\hline
Name     &  Type  & 
\multicolumn{1}{c}{$V_h$} & 
\multicolumn{1}{c}{$V_\mathrm{LG}$} &  
\multicolumn{1}{c}{D$_\mathrm{TRGB}$} & 
$M_B$ & 
TI & 
$\Theta_\mathrm{N253}$  &
$R_\perp$ &
$R_\mathrm{3D}$ \\
         &        & 
\multicolumn{1}{c}{km\,s$^{-1}$} & 
\multicolumn{1}{c}{km\,s$^{-1}$} & 
\multicolumn{1}{c}{Mpc}  & mag      &        & 
deg &
\multicolumn{2}{c}{kpc} \\
\hline
NGC 253                                     & Sc   & $260.6\pm5$ $^a$   & 294 & $3.70\pm0.03$ $^\dagger$    & $-21.15$  &        &     \\
\hspace{4pt} \textbf{Do\,II}                 & dSph &                    &     &                             & $ -6.6\phantom{0}$   & $ 3.2$ &  1.3 &  86 & \\
\hspace{4pt} LV J0055$-$2310                & dIr  & $249.6\pm5$ $^a$   & 288 & $3.62\pm0.18$ $^\ddagger$   & $ -9.81$  & $ 2.2$ &  2.7 & 175 & 191 \\
\hspace{4pt} Scl--MM--Dw1                   & dSph &                    &     & $3.94\pm0.63$ $^\flat$      & $ -9.54$  & $ 1.8$ &  1.1 &  71 & 251 \\
\hspace{4pt} NGC 247                        & Sd   & $153.0\pm5$ $^a$   & 208 & $3.72\pm0.03$ $^\dagger$    & $-18.56$  & $ 1.6$ &  4.5 & 293 & 294 \\
\hspace{4pt} \hspace{4pt} \textbf{Do\,IV}    & dSph &                    &     &                             & $-7.1\phantom{0}$    & $ 2.2$ &  3.6 & 233 & \\
\hspace{4pt} \hspace{4pt} ESO540--032       & Ir   & $227.7\pm0.9$ $^b$ & 285 & $3.63\pm0.05$ $^\dagger$    & $-11.45$  & $ 1.4$ &  5.4 & 349 & 354 \\
\hspace{4pt} \textbf{Do\,III}                & dSph &                    &     &                             & $-8.3\phantom{0}$    & $ 1.4$ &  5.3 & 343 & \\
\hspace{4pt} DDO 6                          & dIr  & $295.3\pm5$ $^a$   & 347 & $3.44\pm0.15$ $^\dagger$    & $-12.45$  & $ 1.3$ &  4.3 & 277 & 373 \\
\hspace{4pt} Sculptor SR                    & dTr  &                    &     &                             & $-10.26$  & $ 1.2$ &  4.0 & 257 & \\
\hspace{4pt} KDG 2                          & dTr  & $223.5\pm2.7$ $^b$ & 290 & $3.56\pm0.07$ $^\dagger$    & $-11.49$  & $ 1.0$ &  7.2 & 463 & 478 \\
\hspace{4pt} Scl--MM--Dw2                   & dSph &                    &     & $3.12\pm0.32$ $^\sharp$     & $-11.03$  & $ 0.7$ &  0.8 &  53 & 582 \\
\hspace{4pt} Sc 22                          & dSph &                    &     & $4.29\pm0.07$ $^\dagger$    & $-10.50$  & $ 0.5$ &  5.4 & 350 & 699 \\
\hspace{4pt} NGC 7793                       & Sd   & $223.8\pm5$ $^a$   & 247 & $3.63\pm0.14$ $^\dagger$    & $-18.34$  & $ 0.2$ & 13.1 & 842 & 838 \\
\hspace{4pt} \hspace{4pt} PGC704814         & Ir   & $270\phantom{.0}\pm89$ $^c$    & 299 & $3.66\pm0.18$ $^\ddagger$   & $-11.61$  & $ 2.1$ & 12.4 & 798 & 795 \\
\hspace{4pt} ESO349--031                    & dIr  & $220.3\pm5$ $^a$   & 229 & $3.21\pm0.06$ $^\dagger$    & $-11.87$  & $ 0.2$ & 12.6 & 802 & 901 \\[4pt]
\hspace{4pt} DDO 226                        & dIr  & $362.1\pm5$ $^a$   & 413 & $4.92\pm0.29$ $^\dagger$    & $-13.63$  & $-0.3$ &  3.2 &     & 1243 \\
\hspace{4pt} UGCA 442                       & Sm   & $266.7\pm5$ $^a$   & 299 & $4.37\pm0.15$ $^\dagger$    & $-14.71$  & $-0.3$ & 15.5 &     & 1274 \\
\hspace{4pt} NGC 625                        & Sm   & $394.9\pm5$ $^a$   & 324 & $4.02\pm0.07$ $^\dagger$    & $-16.50$  & $-0.3$ & 18.9 &     & 1307 \\
\hspace{4pt} NGC 59                         & dEem & $367.6\pm5$ $^a$   & 438 & $4.90\pm0.07$ $^\dagger$    & $-15.57$  & $-0.4$ &  8.3 &     & 1349 \\
\hspace{4pt} PGC 6430                       & Im   & $391.7\pm5$ $^a$   & 306 & $4.57\pm0.13$ $^\dagger$    & $-15.68$  & $-0.7$ & 21.7 &     & 1778 \\
%\hspace{4pt} KK 258                         & dTr  & $\phantom{0}92.0\pm5$ $^d$    & 150 & $2.24\pm0.03$ $^\dagger$ & $-10.51$  & $-0.9$ & 28.4 \\
\hline\hline
\end{tabular}
\newline
\centering
\begin{tabular}{p{0.3\textwidth}p{0.3\textwidth}p{0.3\textwidth}}
$^a$ \citet{2017MNRAS.472.4832W}    &   $^\dagger$ \citet{2009AJ....138..332J}  &   $^\flat$ \citet{2014ApJ...793L...7S} \\
$^b$ \citet{2005AJ....130.2058B}    &   $^\ddagger$ \citet{2021arXiv210211354K} &   $^\sharp$ \citet{2016ApJ...816L...5T} \\
$^c$ \citet{2003astro.ph..6581C}    \\
%$^d$ \citet{2014MNRAS.443.1281K}    &   $^\sharp$ \citet{2016ApJ...816L...5T} \\
\end{tabular}
\end{table*}

The `tidal index' \citep{1999IAUS..186..109K,2013AJ....145..101K} gives us a good proxy of group membership.
It characterizes the degree of the tidal impact from the main disturber (MD) ${\rm TI} \propto \log(L_\mathrm{MD}/R_\mathrm{MD}) + C$,
were $L_\mathrm{MD}$ is a luminosity of the most influential neighbor and $R_\mathrm{MD}$ is its spatial separation from the galaxy under consideration.
The constant $C$ is chosen so that ${\rm TI}=0$ at the `zero-velocity surface' of the main disturber.
Thus, ${\rm TI}>0$ indicates that the galaxy is physically bound to its main disturber, 
while ${\rm TI}<0$ means that the object belongs to the field.
We summarize information on galaxies in the vicinity of NGC\,253 in the Table~\ref{tab:n253group}.
%The columns contains: 
%1) galaxy name;
%2) morphological type;
%3) $V_h$ -- heliocentric velocity in km\,s$^{-1}$ with its error;
%4) $V_\mathrm{LG}$ -- radial velocity with respect to the Local Group centroid \citep{1996AJ....111..794K};
%5) D$_\mathrm{TRGB}$ -- TRGB-distance in Mpc with a corresponding error;
%6) $M_B$ -- $B$-band absolute magnitude of the galaxy;
%7) TI -- `tidal index' indicating the value of the tidal influence from the main disturber \citep{1999IAUS..186..109K,2013AJ....145..101K};
%8) $\Theta_\mathrm{N253}$ -- projected separation from NGC\,253 in degrees;
%9) $R_\perp$ -- projected distance to NGC\,253 in kpc;
%10) $R_\mathrm{3D}$ -- spatial distance to NGC\,253 in kpc.
Morphological type, absolute magnitude, `tidal index' are given according to the current state of the Local Volume database\footnote{\url{https://www.sao.ru/lv/lvgdb/}} \citep{2012AstBu..67..115K}.
Newly found galaxies are marked in bold.
The hierarchy in the group is shown by intent in the galaxy name and
the galaxies are sorted according to the `tidal index' with respect to their main disturber.
Data sources on velocities and distances are indicated as footnotes in the table.
It is necessary to note that the velocity of NGC\,253, $V_h=261$~km\,s$^{-1}$, 
obtained in the deep Parkes \ion{H}{i} survey \citep{2017MNRAS.472.4832W} differs significantly from other measurements.
For instance, HIPASS Bright Galaxy Catalog \citep{2004AJ....128...16K} gives a systemic velocity $V_h=243\pm2$~km\,s$^{-1}$.
\citet{2015MNRAS.450.3935L} measure the velocity of the kinematical center of the galaxy of $V_h=238\pm4$~km\,s$^{-1}$ 
using \ion{H}{i} radio-interferometric observations with the Karoo Array Telescope.
However, to avoid possible systematic, we decided to use the value from the deep Parkes \ion{H}{i} survey \citep{2017MNRAS.472.4832W},
because most velocity measurements around NGC\,253 were taken from this survey.
Note that the choice of one or another velocity of NGC\,253 has practically no effect on the estimate of the total mass of the group.

Most galaxies in the Sculptor filament have precise TRGB distances.
It allows us to map the 3D distribution of matter in this region.
The projections of the spatial distribution of galaxies on the plane of the Local Supercluster and on the perpendicular plane are shown in the Fig.~\ref{fig:n253xyz}.
At the moment there are 12 confirmed members in the group around NGC\,253.
Taking into account three new probable members and Sculptor\,SR for which there is no velocity or distance measurements 
the total population consist of 16 galaxies.
Only 9 of them have know radial velocities, that allows one to estimate the radial velocity dispersion in the NGC\,253 group of 43\,km\,s$^{-1}$.
Most galaxies lie at a projection distance of less than 5.5\degr{} or about 350\,kpc from NGC\,253. In projection,
Do\,II is the third nearest satellite of NGC\,253.
NGC\,247 has high negative peculiar velocity of $-86$\,km\,s$^{-1}$ with respect to NGC\,253.
In fact, Do\,IV together with ESO\,540--032 form a subgroup around NGC\,247.
Do\,III, lying at a projected distance of about 340\,kpc from NGC\,253, 
is one of the peripheral members of the core of the group.
Galaxies NGC\,7793 with its companions PGC\,704814 and ESO\,349--031 are located near the border of the system.
Obviously, these galaxies are just falling into the group center and, 
therefore, they cannot be used for the virial mass estimation.
Other galaxies beyond 1\,Mpc from NGC\,253 participate in a general Hubble expansion,
which allows \citet{2003A&A...404...93K} to determine the zero-velocity radius of the group to be 0.7\,Mpc.

Starburst galaxy NGC\,253 dominates the group.
Its absolute magnitude is 2.6\,mag less (in other words, 11 times brighter) than the second brightest galaxy NGC\,247.
It allows us to use the virial theorem and the projected mass estimator in the simplest form of the test particles around a massive body \citep{1981ApJ...244..805B}.
Using the first five satellites of NGC\,253 with known velocities (LV\,J0055$-$2310 + NGC\,247 + ESO\,540--032 + DDO\,6 + KDG\,2), 
we estimate the virial mass of the group to be $M_\mathrm{vir}=6.4\times10^{11}$\,$M_\sun$.
The projected mass estimator, assuming isotropic orbits distribution $\langle e^2 \rangle = \frac{1}{2}$, 
gives a similar value of $M_\mathrm{pm}=7.1\times10^{11}$\,$M_\sun$.
As we noticed before, these values are almost independent of the adopted velocity of NGC\,253.
In the case of a heliocentric radial velocity of $V_h=243\pm2$ \citep{2004AJ....128...16K}, the corresponding values are
$M_\mathrm{vir}=6.2\times10^{11}$ and $M_\mathrm{pm}=6.8\times10^{11}$\,$M_\sun$.
Taking into account that KDG\,2 lying at a distance more than 460\,kpc is most likely located outside the virial radius and excluding it from the analysis, 
for the 4 remaining satellites we obtain $M_\mathrm{vir}=7.3\times10^{11}$ and $M_\mathrm{pm}=8.8\times10^{11}$\,$M_\sun$.
So, adopting the total mass of the NGC\,253 group is about $8\times10^{11}$\,$M_\sun$,
one can estimate the radius of the virialized region of $R_{200}=186$\,kpc,
which corresponds to a sphere whose density is 200 times greater than the critical one.
According to an exact analytical solution \citep{2019MNRAS.490L..38B} the corresponding radius of the zero velocity surface is 706\,kpc.
This value is in excellent agreement with the direct measurement of the zero velocity radius of 0.7\,Mpc by \citet{2003A&A...404...93K}.

%First of all, this galaxy appears in my catalog of groups
%https://ui.adsabs.harvard.edu/abs/2011MNRAS.412.2498M/abstract  
%as a main member of the group [MK2011]NGC0253
%http://galaxies1.univ-lyon1.fr/ledacat.cgi?o=[MK2011]NGC0253
%I have recalculated it several times the last version contains 4 galaxies with the main member of  NGC\,253.
%NGC0253 + IC1574 + NGC0247 + ESO540-031
%Also previous iterations include NGC0059 in this group. So, probably it is distant member of the group.
%My catalog of groups has an advantage that it uses physically motivated restrictions on the clusterization. So, I believe it is not too contaminated by false members. However, this method based only on radial velocities of galaxies and the data are 9 years old.

%The catalog of the galaxies in the Local Group allows to associate the members using so called a tidal index, TI.
%We investigated the suits of dwarfs around NGC\,253 (Karachentsev, Kaisina \& Makarov 2014). 2014AJ....147...13K
%The current version of our database contains 10 satellites of NGC\,253 taking into account that TI>0 separates confined objects from the isolated ones:
%LVJ0055-2310 + Scl-MM-Dw1 + NGC0247 + DDO006 + ScuSR + KDG002 + Scl-MM-Dw2 + Sc22 + NGC7793 + ESO349-031.
%The outskirt of the group with $0<TI<-0.5$ consists of 4 galaxies: 
%DDO226 + NGC0625 + UGCA442 + NGC0059.

%\begin{figure}
%    \centering
%    \includegraphics[width=0.49\textwidth]{ngc253_angle_cz.eps}
%    \caption{Angle -- redshift diagram.}
%    \label{fig:n253angle}
%\end{figure}

\subsection{A Plane of Satellite Galaxies?}

\begin{figure}
\centering
\includegraphics[width=0.49\textwidth]{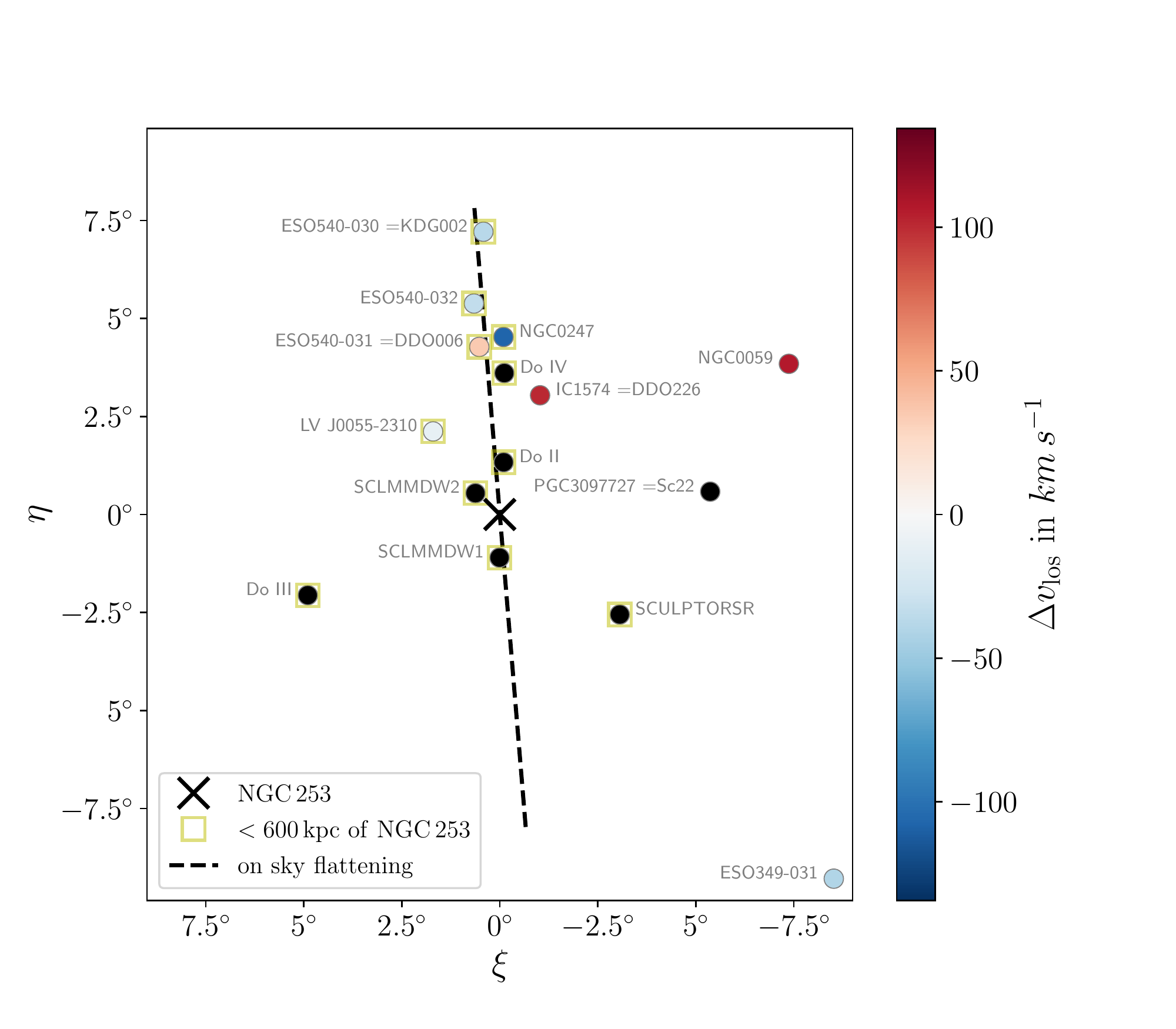}
\caption{
On-sky distribution of the galaxies listed in Table~\ref{tab:n253group} in equatorial coordinates relative to the position of NGC\,253 (black cross). 
The galaxies are color-coded by their line-of-sight velocity component relative to the NGC\,253 velocity. 
Galaxies without velocity measurements are plotted in black. 
Galaxies (potentially) within 600\,kpc of NGC\,253 are marked with yellow boxes. 
The dashed black line indicates the on-sky orientation of the flattening of these objects.
}
\label{fig:onsky}
\end{figure}

\begin{figure*}
    \centering
    \includegraphics[width=0.24\textwidth]{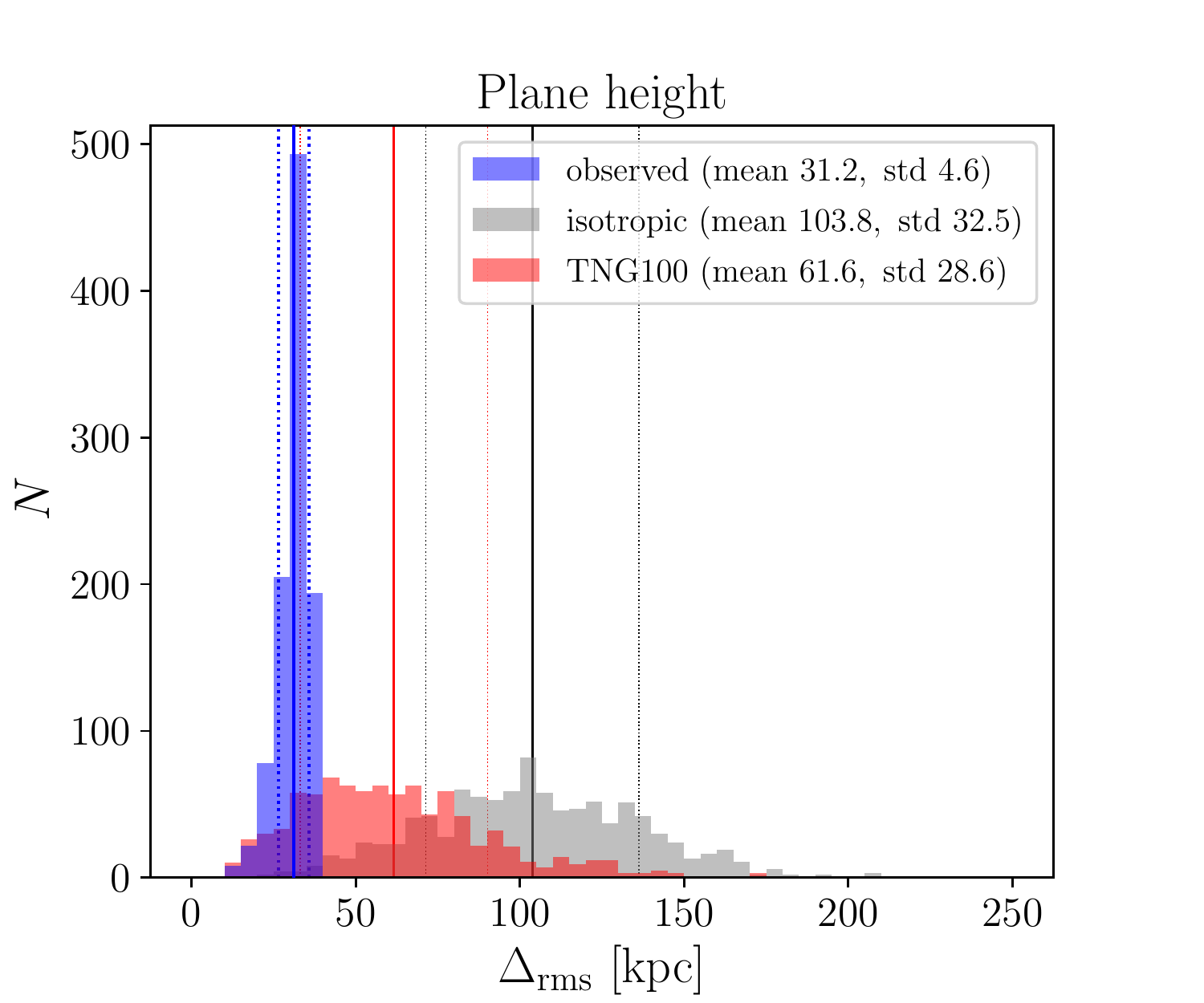}
    \includegraphics[width=0.24\textwidth]{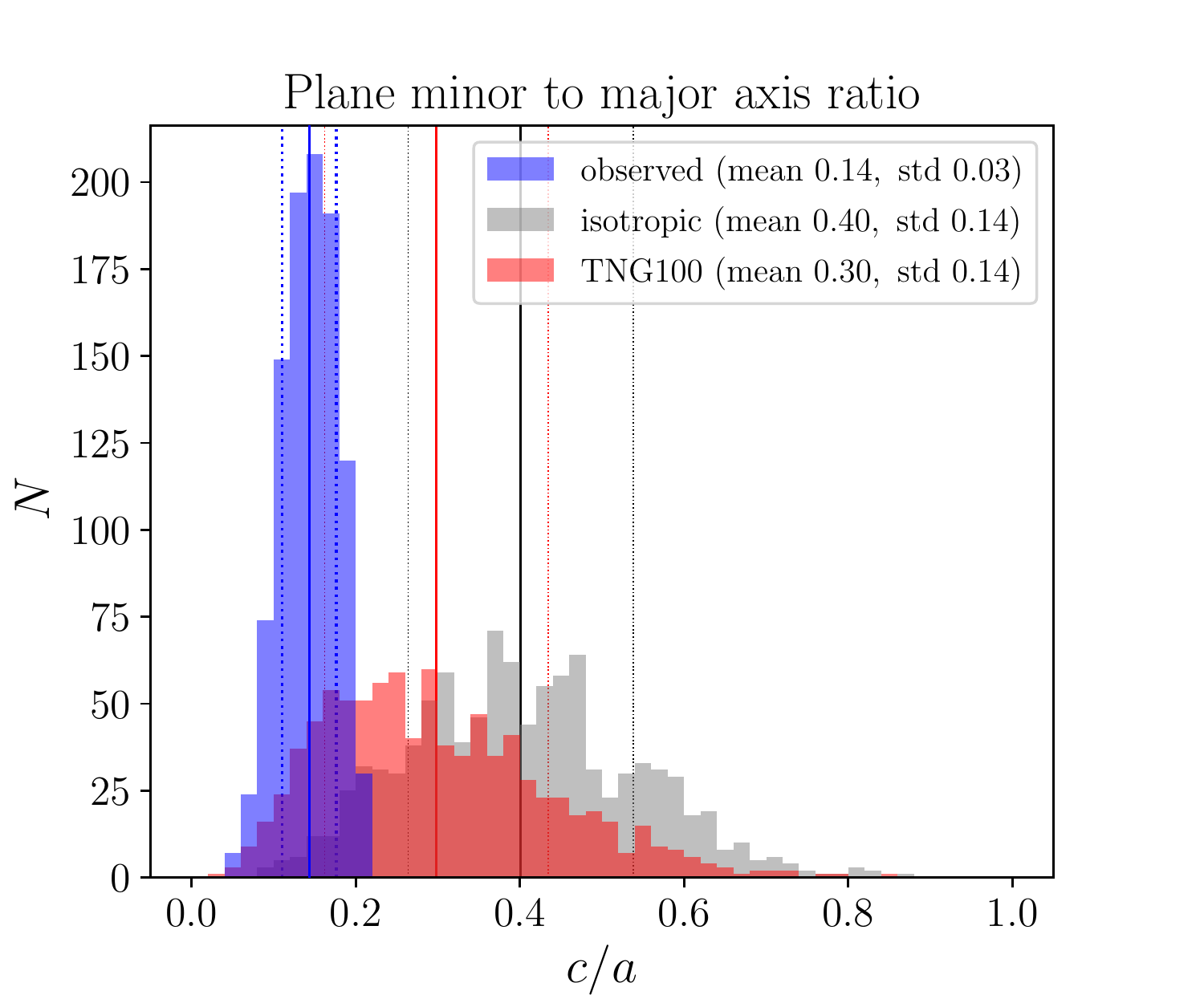}
    \includegraphics[width=0.24\textwidth]{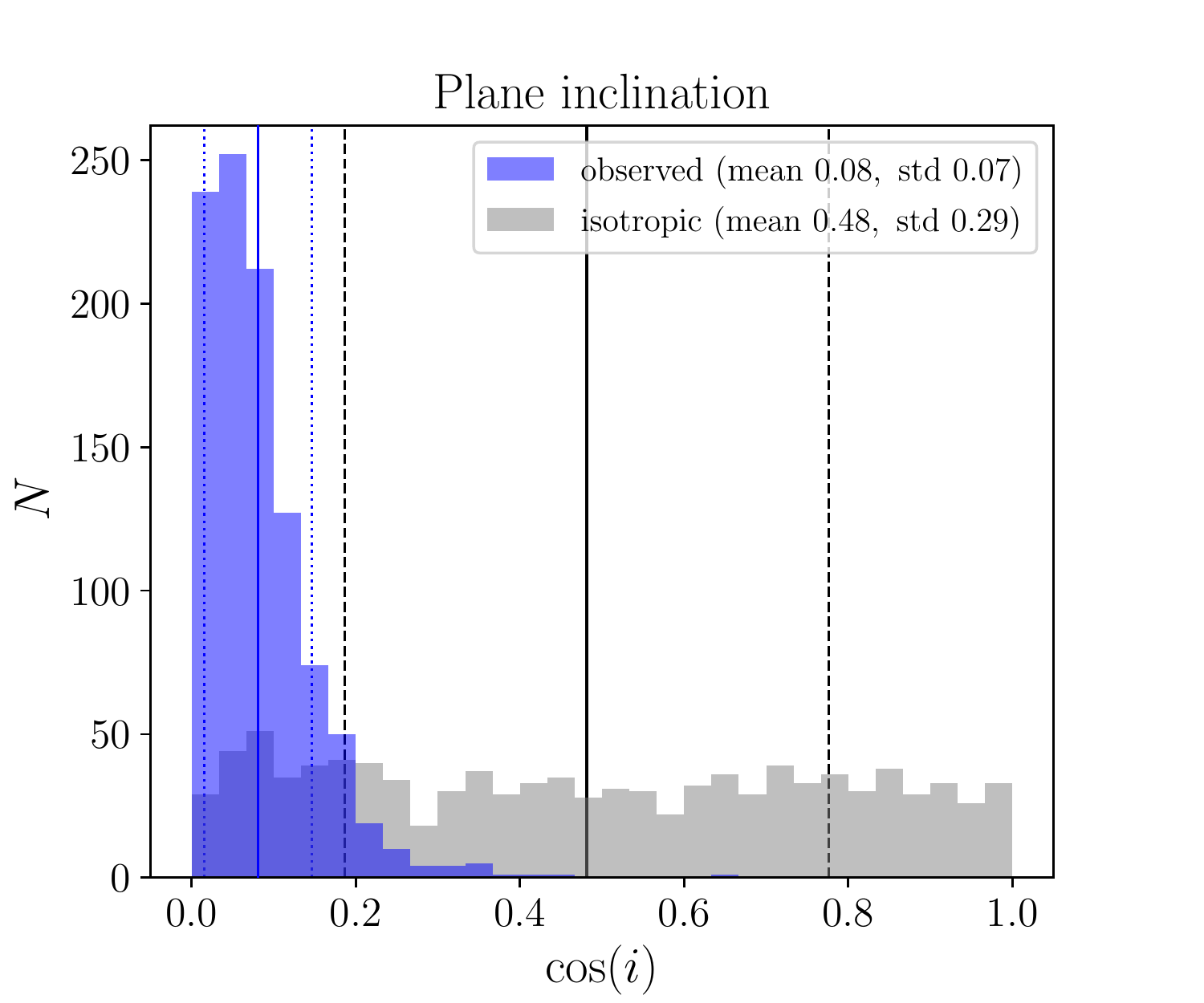}
    \includegraphics[width=0.24\textwidth]{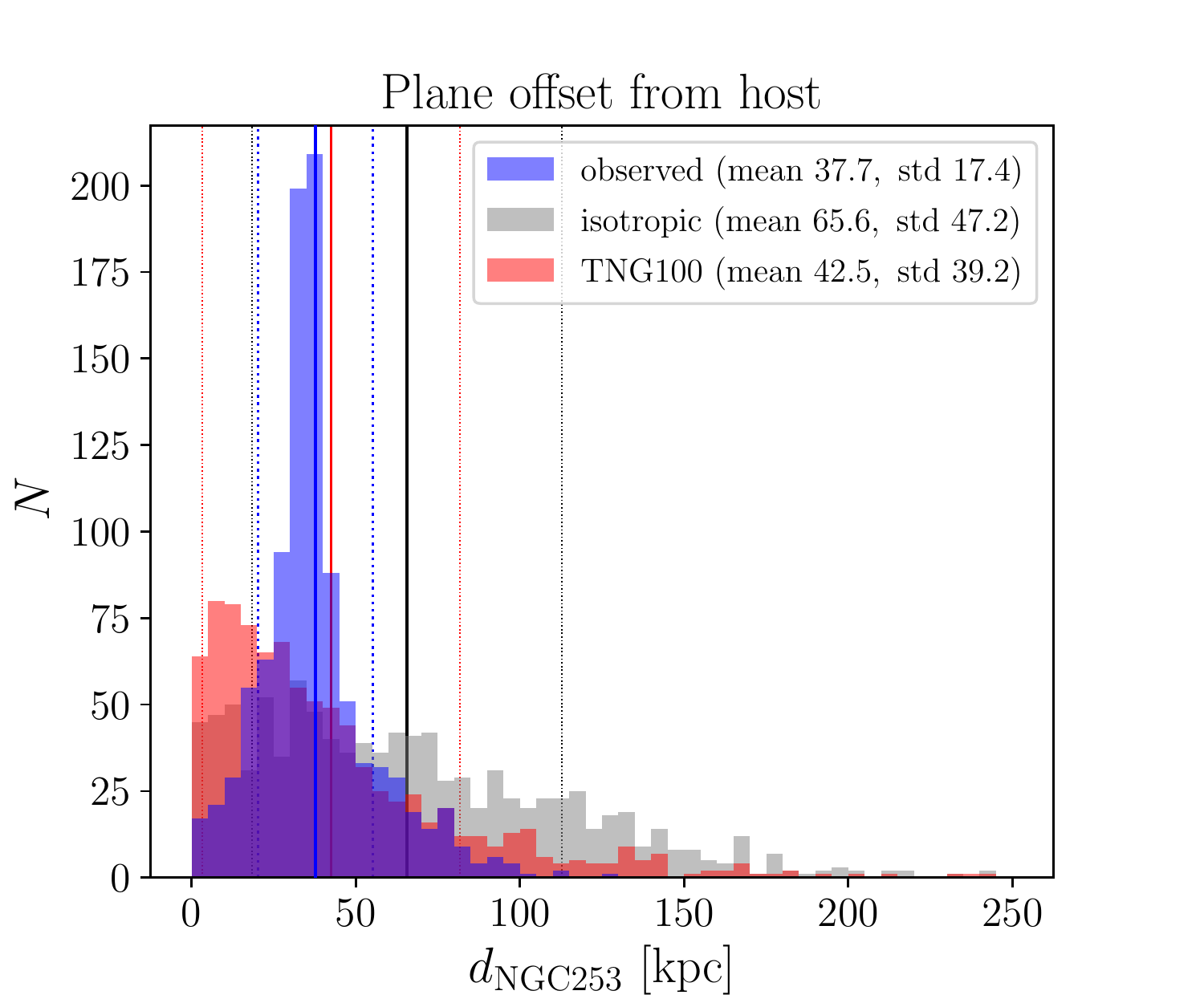}
    \caption{Best-fit parameters for the ToI fits to the spatial distributions of galaxies within 600\,kpc of NGC\,253. Shown are, from left to right: the absolute rms plane height $\Delta_\mathrm{rms}$, the relative minor-to-major axis ratio $c/a$, the inclination $i$\ of the best-fit plane with the line-of-sight to NGC\,253, and the perpendicular offset $d_\mathrm{NGC\,253}$\ of the best-fit plane from the host galaxy NGC\,253 (which was not included in the plane fit). Each panel also gives the mean and standard deviations of the shown distributions. The realizations drawing from the observed galaxy distances (blue) are much more flattened (by both measures) than the randomized samples (grey) drawn from isotropic distributions around NGC\,253's position but following the same radial distances. Satellite systems drawn from the Illustris TNG 100 simulation (red) are more flattened than isotropic distributions in both absolute plane height and axis ratio, but the observed system falls onto the more flattened tail of their distribution.}
    \label{fig:planefit}
\end{figure*}

\begin{figure}
    \centering
    \includegraphics[width=0.49\textwidth]{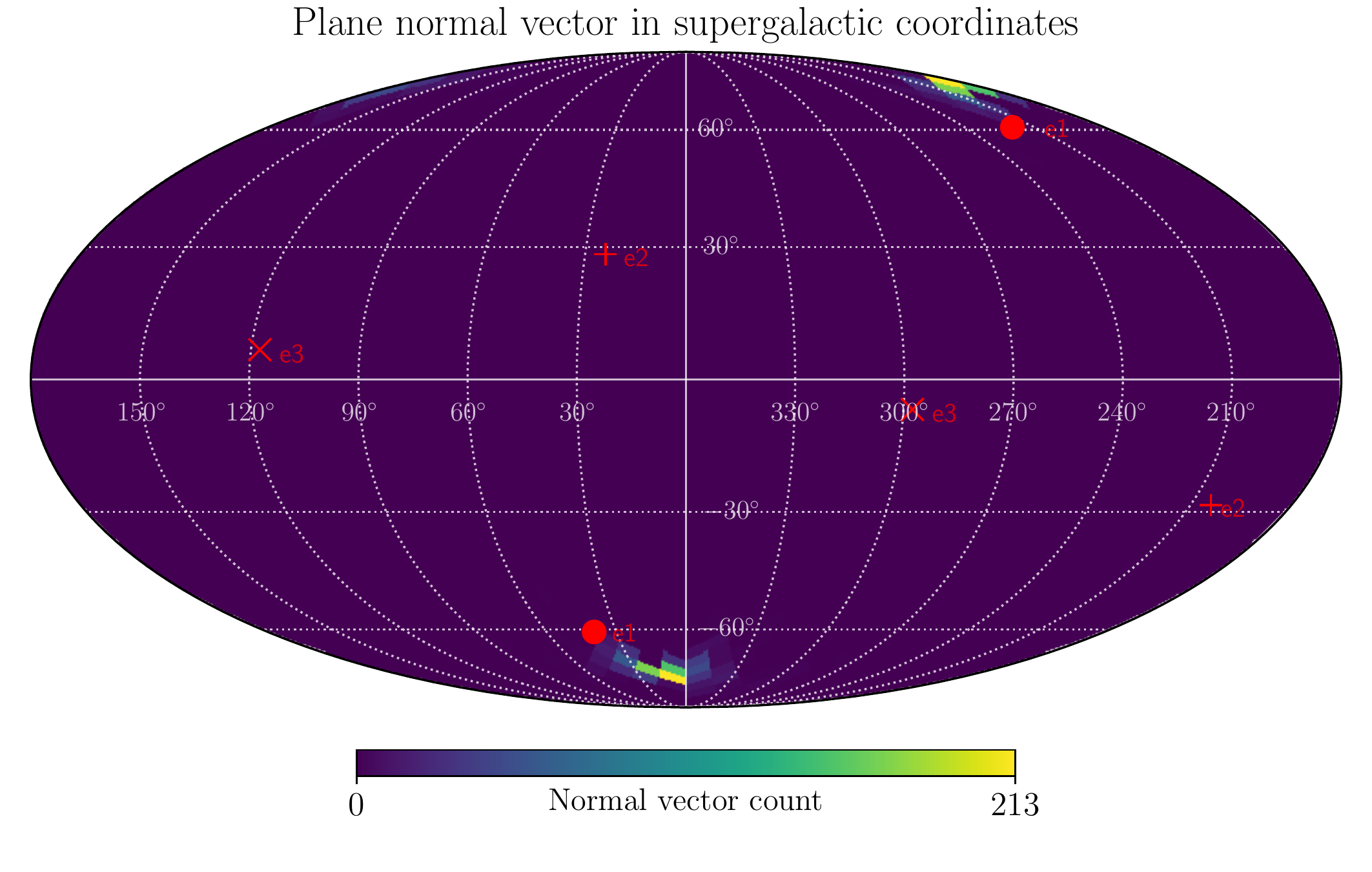}
    \caption{Density distribution of the direction of plane-normal vectors (minor axis of 3D galaxy distribution) of the galaxies with measured distances within 600\,kpc around NGC\,253, for 1000 realizations drawing from the distance errors. Shown are Supergalactic Coordinates, indicating that the best-fit plane aligns closely with the Supergalactic Plane. The red symbols indicate the eigenvectors of the tidal tensor at the position of NGC\,253, which describe the principal directions of the surrounding cosmic web. The best-fit plane is well aligned with $\vec{e}_1$, the axis along which material in the cosmic web is compressed fastest.}
    \label{fig:planenormal}
\end{figure}

The presence of spatially flattened and velocity-correlated satellite galaxy systems around the Milky Way, M\,31, and Centaurus\,A motivates the question: Is there evidence for a similar alignment among the satellites of NGC\,253?

Fig.~\ref{fig:onsky} shows the on-sky distribution of the galaxies listed in Table~\ref{tab:n253group}, with the galaxies color-coded by their line-of-sight velocity component relative to the NGC\,253 velocity. It is apparent from this figure that most of the known objects in the vicinity of NGC\,253 align in an extremely narrow structure to the north. Of the 11 galaxies that potentially are within 600\,kpc of NGC\,253 (indicated by yellow boxes), only Do\,III and Sculptor\,SR are significantly offset from this orientation. Neither of these two have measured distances, so they could well be fore- or background objects outside of the system.

This asymmetric distribution of galaxies with respect to the NGC\,253 can not be explained simply by the observational biases.
The group is located near the South Galactic Pole, where the Galactic extinction is insignificant.
A simple count of galaxies in a 30-degree cone around NGC\,253 using the HyperLeda database \citep{2014A&A...570A..13M} shows that the redshift data is 97 per cent complete down to $B_T=15.5$ and 86 per cent complete down to $B_T=16.6$~mag.
Thus we believe that we know velocities of all possible satellites in the group with $M_B\lesssim-12.3$~mag.
Table~\ref{tab:n253group} confirms this estimate. 
The only satellites without redshift measurements are dSph galaxies fainter than $M_B=-11$~mag.
The hunt for more faint galaxies of extremely low surface brightness without gas is extremely difficult task,
and our knowledge of the faint end of the galaxy luminosity function is highly incomplete below $M_B\sim-11$~mag.
However, the brighter galaxies also show the similar spatial distribution, 
that appears to reflect the real structure of the group.

Of the eleven galaxies possibly within 600\,kpc of NGC\,253, eight are in the North and only three are in the South, and the asymmetry increases to six in the North and only one in the South if the four galaxies without distance measurements are excluded. The distribution of known galaxies around NGC\,253 is thus strongly lopsided. While the asymmetry appears to be quite extreme in this case, lopsided distributions of galaxies around a host appear to be rather common: they have been found for the distribution of satellite galaxies around M31 \citep{2013ApJ...766..120C}, in statistical samples of both paired \citep{2016ApJ...830..121L} and isolated galaxies \citep{2020ApJ...898L..15B}, and even in cosmological simulations \citep{2017ApJ...850..132P, 2021arXiv210412787W}.

It is intriguing that of the five galaxies with line-of-sight velocities, all of which align along the flattened structure to the North of NGC\,253, four are blue- and only one is redshifted relative to NGC\,253. This is well consistent with an interpretation as a mostly rotating satellite plane like those seen in M\,31 and Centaurus\,A. In fact, a fraction of 80 per cent correlated velocities is very similar to the fractions of correlated velocities reported for M\,31 (13 out of 15, or 86 per cent) by \citet{Ibata2013}, and for Centaurus\,A (21 out of 28, or 75 per cent) by \citet{Mueller2021}. However, due to the small number of objects, the degree of correlation for the NGC\,253 system is not significant. Finding a velocity coherence of at least 4 out of 5 happens in 37.5 per cent of all cases. A larger number of galaxies with measured velocities will thus be required for a statistically meaningful analysis, demonstrating an urgent need for spectroscopic follow-up of candidate dwarf galaxies in this region.

To study the three-dimensional distribution of the known satellites of NGC\,253, we employ the common tensor-of-inertia (ToI) fitting method \citep[see e.g.][]{Pawlowski2015}. We account for the uncertainties in the distance measurements by generating 1000 realizations of the observed system, drawing galaxy distances from normal distributions that are centered on their most-likely value and have a $1\sigma$-width of the errors, as compiled in Table~\ref{tab:n253group}. We consider all galaxies within 600\,kpc, excluding the four candidates without measured distances (Do\,II, III, IV and Sculptor\,SR). The results of our plane fits are compiled in Fig.~\ref{fig:planefit}. As expected from the flattened on-sky distribution, we find an extremely narrow spatial alignment: the root-mean-square (rms) height from the best-fit plane is only $\Delta_\mathrm{rms} = 31 \pm 5$\,kpc, and the minor-to-major axis ratio is $c/a = 0.14 \pm 0.03$. These values are comparable to those of the satellite planes around the Milky Way ($\Delta_\mathrm{rms} = 20\ \mathrm{to}\ 30$\,kpc, $c/a \approx 0.2$) and M\,31 ($\Delta_\mathrm{rms} < 14.1$\,kpc, $c/a \approx 0.1$). The normal direction to the best-fit plane is shown in Fig.~\ref{fig:planenormal}. It points to Supergalactic Coordinates $(l_\mathrm{SG}, b_\mathrm{SG}) = (12^\circ, -73^\circ)$, and varies by only $4^\circ$\ around this direction. The plane has an inclination of $i = 85^\circ$, and is thus seen very close to edge-on -- as is the case for the M\,31 and Centaurus\,A satellite structures. The good alignment with the Supergalactic plane could be related to the preference of flattened satellite galaxy structures in the Local Universe to align with the direction of greatest collapse of their surrounding larger-scale structure \citep{Libeskind2019}. This could also explain the preference to see these structures edge-on (since both the Milky Way and the nearby hosts lie in a common sheet-like structure). Indeed, as Fig.~\ref{fig:planenormal} shows the plane normal is closely aligned with the $\vec{e}_1$-direction, to within $(16.6 \pm 5.4)^\circ$, where $\vec{e}_1$\ is the eigenvectors of the tidal tensor at the position of NGC\,253 which corresponds to the axis along which material in the cosmic web is compressed fastest. NGC\,253 thus further confirms the finding in \citet{2018MNRAS.473.1195L} that some flattened dwarf galaxy structures in the Local Universe preferentially align with the $\vec{e}_1$-direction at their respective positions, and thus appear to be related to the larger-scale cosmic web.

To assess the significance of the found flattening compared to completely random, uncorrelated distributions, we repeat the plane-fitting on randomized systems. These are drawn from an isotropic distribution centered on NGC\,253, but follow the exact same radial distances from the host as our observed realizations (grey histograms in the figure). The isotropic distributions typically results in much wider plane fits, and the flattening measured for the observed system are off by $2\sigma$ for both the absolute plane height and the relative axis ratios. Thus, while not highly significant -- as expected given the small number of objects -- it is still an intriguing alignment that warrants further study of this system's phase-space correlation, in particular via distance measurements of other satellite galaxy candidates to assess their association with NGC\,253.

While a detailed comparison to cosmological expectations is beyond the scope of the present work, we nevertheless illustrate the expected flattening of satellite galaxy systems in $\Lambda$CDM by comparing to systems in the Illustris TNG-100 hydrodynamical cosmological simulation \citep{2018MNRAS.475..676S, 2019ComAC...6....2N}. As analogs, we select host galaxies with virial masses in the range of 0.6 to $1.0 \times 10^{12}\,M_\sun$, which are required to be  sufficiently isolated by rejecting all hosts that have another galaxy with a virial mass exceeding $0.5 \times 10^{12}\,M_\sun$ within 1.2\,Mpc. The galaxies within a volume of 600\,kpc around each host are ranked based on their brightness (if they contain stars), followed by a ranking by mass (if they are dark). The flattening of the seven top-ranked galaxies (excluding the host) is measured, and the resulting distribution of absolute plane height, relative axis ratio, and best-fit plane's offset from the host are plotted in red in Fig. \ref{fig:planefit}.

As expected, the $\Lambda$CDM satellite systems are more flattened than systems drawn from isotropy (especially because the considered volume exceeds the virial volume of the simulated hosts). However, the observed distribution of galaxies around NGC\,253 remains on the more flattened tail of the distribution: only $10.1^{+7.2}_{-5.3}$\ and $12.5^{+6.8}_{-4.6}$\  per cent of the simulated systems are more flattened than the observed system in $c/a$\ and $\Delta_\mathrm{rms}$, respectively (where the error bars are based on the standard deviations obtained from the Monte-Carlo sampling of the observed distribution). The flattened distribution of the presently confirmed galaxies in the vicinity of NGC\,253 thus does not rise to the same degree of tension reported for the better-studied satellite galaxy planes around the Milky Way, M31, and Centaurus A.

If, however, only one additional satellite would be confirmed, without changing the overall measured flattening of the system, then the fractions reported above would already drop by about a factor of two, to $5.4^{+4.9}_{-2.9}$\ and $7.1^{+5.0}_{-3.6}$\ per cent of simulated systems being more flattened in $c/a$\ and $\Delta_\mathrm{rms}$, respectively. This demonstrated that it will require a more complete census of galaxies around NGC\,253 via distance measurements of the current candidates, as well as a more complete sample of spectroscopic velocity measurements, to clarify whether the system is indeed exceptional, or rather typical, compared to expectations derived from cosmological simulations.

If we assume that the galaxies (including our new three discoveries) for which no distance measurement exist are at the distance of NGC\,253, with a broad distance uncertainty of $\pm 200$\,kpc, our plane fits become less extreme. 
The broad assumed distance uncertainties also introduce a lot of scatter in the obtained best-fit parameters. 
While the best-fit orientation changes only mildly, the plane width are then dominated by the two spatial outliers Do\,III and Sculptor\,SR.

While one can interpret the spatial alignment as a potential plane of satellite galaxies, its extent of 600\,kpc exceeds the virial volume of NGC\,253. This makes the structure more similar to the larger-scale planes of dwarf galaxies discovered in the LG, which have diameters of 1--2\,Mpc \citep{Pawlowski2013}. 
However, even the 6 galaxies (assuming those without distance measurements share the same distance as NGC\,253) within 300\,kpc are roughly aligned in the same direction as the objects on larger scales. 
Another interpretation of the larger structure is that these galaxies trace a cosmic filament. 
Preferential accretion of dwarf galaxies along filaments was proposed as a possible origin of flattened satellite galaxy structures \citep{Lovell2011, Libeskind2011}. 
While the alignment of the flattened galaxy distribution with the direction of fastest collapse of the surrounding cosmic web can be seen as support of such an interpretation, the observed distribution appears much more narrow than a typical filament in cosmological simulations, and the resulting satellite galaxy systems in simulations are typically not highly anisotropic \citep{Pawlowski2012, Shao2018}. 
However, since the effects of accretion along filament on the formation of satellite alignments have thus far mostly been studied in cosmological simulations, NGC\,253 promises to be a unique observational test case of the filamentary accretion scenario: a potential satellite plane in the making by the cosmic web.

\section{Conclusions}

In this study we report the discovery of three dwarf spheroidal galaxies, named Do\,II, Do\,III and Do\,IV, in the vecinity of the bright late-type spiral NGC\,253 galaxy by means of a visual inspection of the %. We performed new NGC\,253 satellite search using coadded
images taken by the DES. For our analysis, the images
were processed using the \textsc{legacypipe} software from the DESI Legacy imaging surveys.
%Each image was astrometrically calibrated to Gaia-DR2 and photometrically calibrated
%to the Pan-STARRS PS1 survey. %The resulting image includes approximately 8 exposures
%in each of $g$, $r$, and $z$-bands.
%As a result of our search, three new dwarf galaxies have been discovered in the vicinity of bright late-type spiral NGC\,253. We named them Do\,II, Do\,III and Do\,IV. 
We used the \textsc{galfit} software for deriving their photometric and structural properties. The physical parameters for Do\,II, Do\,III and Do\, IV were calculated
by adopting a distance modulus of $27.84\pm0.02$\,mag for NGC\,253 \citep{2009AJ....138..332J}.
The resulting total absolute magnitudes of the dwarfs transformed to the $V$-band fall
in the range from about $-7$ to about $-9$\,mag, which are typical for satellite
dwarf galaxies in the Local Universe. 
The central surface brightness tend to be extremely low for all the discovered dwarfs and fall roughly in the range of 25--26\,mag\,arcsec$^{-2}$ in $g$-band. 

So far there are 12 confirmed members in the group around NGC\,253. 
Taking into account these three possible new  members and the Sculptor\,SR dwarf galaxy, the total population consists of 16 galaxies. 
The radial velocity dispersion in the NGC\,253 group estimated using the nine members
with known radial velocities is 43~km\,s$^{-1}$. 
Most galaxies lie at a projected distance of less than 5.5\degr{} or about 350\,kpc from NGC\,253. 
Do\,II is the third nearest in projection satellite of NGC\,253. 
The late spiral galaxy NGC\,247 has high negative peculiar velocity of $-86$\,km\,s$^{-1}$ with respect to NGC\,253.
In fact, Do\,IV together with the dwarf irregular galaxy ESO\,540--032 forms a subgroup around NGC\,247. 
Do\,III, lying at a distance of about 340\,kpc from NGC\,253, is one of the peripheral members of the core of the group. 

We have used the virial theorem and the projected mass estimator to determine the galaxy group mass. 
We estimated the total mass of the NGC\,253 group to be about $8\times10^{11}M_{\sun}$. 
Therefore, the respective estimated radius of the virialized region is $R_{200} = 186$\,kpc, 
and the corresponding radius of the zero velocity surface is 706\,kpc.

We also discuss the issue of the possible existence of a spatially
flattened and velocity-correlated satellite galaxy system around NGC\,253. It is apparent from our analysis that most of the known objects in the vicinity of NGC\,253 are
aligned in an extremely narrow structure to the North side of the galaxy. Of the 12 galaxies that
potentially are within 600\,kpc from NGC\,253, only Do\,III and Sculptor\,SR are
significantly offset from this orientation. 

To study the three-dimensional distribution
of the known satellites of NGC\,253, we have employed the common tensor-of-inertia
fitting method. As expected from a flattened on-sky distribution, we have found
an extremely narrow spatial alignment: the root-mean-square (rms) height from the best-fit plane is only $\Delta_\mathrm{rms} = 31 \pm 5$\,kpc, and the minor-to-major axis ratio is $c/a = 0.14 \pm 0.03$. These values are comparable to those of the satellite planes found around the Milky Way %($\Delta_\mathrm{rms} = 20\ \mathrm{to}\ 30$\,kpc, $c/a \approx 0.2$) 
and M\,31. %($\Delta_\mathrm{rms} < 14.1$\,kpc, $c/a \approx 0.1$). 
While one can interpret this alignment as a potential plane of
satellite galaxies, its extension of 600\,kpc exceeds the virial volume
of NGC\,253. This makes the structure more similar to the larger-scale planes of dwarf galaxies discovered in the LG, which have diameters of 1--2\,Mpc  \citep{Pawlowski2013}.

The first results of our survey of dwarf galaxies around NGC\,253 or the recent discovery of a possible very faint satellite of M\,33 \citep{2021arXiv210403859M} demonstrated the suitability of DESI Legacy surveys imaging for discovering  diffuse satellites of nearby spiral galaxies by means of the visual inspection of the images.  Because of the low surface brightness of the three dwarf galaxies reported here, their association with NGC\,253 can be only confirmed with TRGB estimates in CMDs from deeper HST data and ground-based 8-meter telescopes or, alternatively, radial velocities obtained with state-of-the-art spectroscopic instruments \citep[e.g.,][]{2021A&A...645L...5M}. 

 %  \begin{enumerate}
%      \item The conditions for the stability of static, radiative
%         layers in gas spheres, as described by Baker's (\citeyear{baker})
%         standard one-zone model, can be expressed as stability
%         equations of state. These stability equations of state depend
 %        only on the local thermodynamic state of the layer.
%      \item If the constitutive relations -- equations of state and
%         Rosseland mean opacities -- are specified, the stability
%         equations of state can be evaluated without specifying
%         properties of the layer.
%      \item For solar composition gas the $\kappa$-mechanism is
%         working in the regions of the ice and dust features
%         in the opacities, the $\mathrm{H}_2$ dissociation and the
%         combined H, first He ionization zone, as
%         indicated by vibrational instability. These regions
%         of instability are much larger in extent and degree of
%         instability than the second He ionization zone
%         that drives the Cephe{\"\i}d pulsations.
%   \end{enumerate}

\begin{acknowledgements}

We thank the referee for giving constructive comments which substantially
helped improving this paper.
We also thank Noam Libeskind, Martha Haynes and Lister Staveley-Smith for useful  comments. 
DMD acknowledges financial support from the Talentia Senior Program (through the incentive ASE-136) from Secretar\'\i a General de  Universidades, Investigaci\'{o}n y Tecnolog\'\i a, de la Junta de Andaluc\'\i a. 
DMD also acknowledges funding from the State Agency for Research of the Spanish MCIU through the ``Center of Excellence Severo Ochoa" award to the Instituto de Astrof{\'i}sica de Andaluc{\'i}a (SEV-2017-0709). 
MSP thanks the Klaus Tschira Stiftung and German Scholars Organization for support via a KT Boost Fund. JR acknowledge financial support from the grants AYA2015-65973-C3-1-R and RTI2018-096228-B-C31 (MINECO/FEDER, UE) as well as support from the State Research Agency (AEI-MCINN) of the Spanish Ministry of Science and Innovation under the grant "The structure and evolution of galaxies and their central regions" with reference PID2019-105602GB-I00/10.13039/501100011033.
We acknowledge the usage of the HyperLeda database\footnote{\url{http://leda.univ-lyon1.fr}} \citep{2014A&A...570A..13M}.

This project used public archival data from the Dark Energy Survey. Funding for the DES Projects has been provided by the U.S. Department of Energy, the U.S. National Science Foundation, the Ministry of Science and Education of Spain, the Science and Technology FacilitiesCouncil of the United Kingdom, the Higher Education Funding Council for England, the National Center for Supercomputing Applications at the University of Illinois at Urbana-Champaign, the Kavli Institute of Cosmological Physics at the University of Chicago, the Center for Cosmology and Astro-Particle Physics at the Ohio State University, the Mitchell Institute for Fundamental Physics and Astronomy at Texas A\&M University, Financiadora de Estudos e Projetos, Funda{\c c}{\~a}o Carlos Chagas Filho de Amparo {\`a} Pesquisa do Estado do Rio de Janeiro, Conselho Nacional de Desenvolvimento Cient{\'i}fico e Tecnol{\'o}gico and the Minist{\'e}rio da Ci{\^e}ncia, Tecnologia e Inova{\c c}{\~a}o, the Deutsche Forschungsgemeinschaft, and the Collaborating Institutions in the Dark Energy Survey.

The Collaborating Institutions are Argonne National Laboratory, the University of California at Santa Cruz, the University of Cambridge, Centro de Investigaciones Energ{\'e}ticas, Medioambientales y Tecnol{\'o}gicas-Madrid, the University of Chicago, University College London, the DES-Brazil Consortium, the University of Edinburgh, the Eidgen{\"o}ssische Technische Hochschule (ETH) Z{\"u}rich,  Fermi National Accelerator Laboratory, the University of Illinois at Urbana-Champaign, the Institut de Ci{\`e}ncies de l'Espai (IEEC/CSIC), the Institut de F{\'i}sica d'Altes Energies, Lawrence Berkeley National Laboratory, the Ludwig-Maximilians Universit{\"a}t M{\"u}nchen and the associated Excellence Cluster Universe, the University of Michigan, the National Optical Astronomy Observatory, the University of Nottingham, The Ohio State University, the OzDES Membership Consortium, the University of Pennsylvania, the University of Portsmouth, SLAC National Accelerator Laboratory, Stanford University, the University of Sussex, and Texas A\&M University.

Based in part on observations at Cerro Tololo Inter-American Observatory, National Optical Astronomy Observatory, which is operated by the Association of Universities for Research in Astronomy (AURA) under a cooperative agreement with the National Science Foundation.
\end{acknowledgements}

\bibliographystyle{aa} % style aa.bst
\bibliography{ref} % your references Yourfile.bib

\end{document}